\documentclass[reprint, aps, pra, amsmath, amssymb]{revtex4-2}

\usepackage{amsmath}
\usepackage{microtype}
\usepackage{bm}
\usepackage{graphicx}
\usepackage{titlesec}
\usepackage{amssymb}
\usepackage{dcolumn}
\usepackage{hyperref}

\begin{document}
\title{Intrinsic Berry Curvature on the Higher-order Poincaré Sphere}

\author{Kaijun Zhang}
\author{Boen Yang}
\author{Hailu Luo}
\email{hailuluo@hnu.edu.cn}
\affiliation{Laboratory for Spin Photonics, School of Physics and Electronics, Hunan University, Changsha 410082, China}
\date{\today}

\begin{abstract}
Higer-order geometric phase enable versatile spin-orbit manipulation of structured light. Conventional theories describe this phase via a total-angular-momentum dependent Berry curvature. We develop an intrinsic theory restricted to the effective two-dimensional subspace, where the Berry curvature is independent of the topological charge of the light field. In our framework, the total-angular-momentum difference does not alter the effective monopole strength, whereas it controls the solid angle enclosed by the trajectory of the state evolution, insteadly. Our theory provides a different topological interpretation of geometric phase of the structured light, which unifies the underlying structures of generalized Poincaré Spheres. Furthermore, it does not impose any specific form on the physical operation of the optical element, naturally encomoassing the case of previously reported experiment, and can be applied in arbitrary state evolutions in both classical structured light and quantum spin-orbit systems.
\end{abstract}

\maketitle

The Berry phase, proposed by Berry in 1984 \cite{10.1098/rspa.1984.0023}, arises when a quantum system is adiabatically transported through a cyclic variation of external parameters. A well-known example \cite{10.1103/PhysRevLett.58.1593} is a spin-1/2 particle in a slowly varying magnetic field. In this case, the Hamiltonian depends only the direction of the field. As the field direction evolves, the quantum state follows this change and can be effectively represented as a point moving on Bloch sphere. When the system completes a closed loop on this sphere, the quantum state acquires an additional phase beyond the dynamical contribution. This kind of Berry phase depends only on the geometry of the path in parameter space \cite{10.1103/PhysRevLett.60.2339}.

The essential feature of the preceding example is not specific to quantum spin systems, but rather to the geometric structure of a representation of state, and the mapping from the state space to the paramter space. A similar phenomenon appears in classical optics \cite{10.1002/lpor.202100003}. The polarization state of light can be described by a two-component complex vector, also can be mapped to a point on a sphere, known as the Poincaré sphere (PS) \cite{Poincaré1892, 10.1364/JOSA.44.000634}, after removing the redundant global phase. When polarized light undergoes a sequence of transformations that brings it back to its initial state, the output field acquires an additional phase. This phase depends only on the trajectory of the polarization state on the Poincaré sphere and is known as the Pancharatnam-Berry (PB) phase \cite{Pancharatnam1956, 10.1080/09500348714551321, 10.1016/S0030-4018(99)00473-3, 10.1364/JOSAA.485485}.

A more general class of geometric phase singularities \cite{10.1016/S0370-1573(96)00029-4, 10.1364/OL.24.000430, 10.1103/PhysRevLett.90.203901, 10.1103/PhysRevLett.97.043901} associated with spatial mode structure gives rise to optical oribital angular momentum (OAM) states \cite{10.1103/PhysRevA.45.8185, 10.1088/2040-8978/19/1/013001, 10.1364/AOP.1.000001, 10.1117/1.AP.3.6.064001, 10.1063/5.0097923, 10.1063/5.0194981, 10.1103/PhysRevA.102.031501}. With introduction of the OAM of light, the polarization basis can be generalized to spin-orbit coupled states carrying distinct topological charges, leading to higher-order Poincaré sphere (HOPS) \cite{10.1103/PhysRevLett.107.053601, 10.1364/OL.39.005274, 10.1364/OE.23.026586, 10.1038/nphoton.2016.37}. Consequently, the explicit form of higher-order geometric phase \cite{10.1103/PhysRevLett.108.190401}, which should be associated with Berry curvature on the generalized Poincaré Sphere, are followed with interest. Commonly, Berry curvature on the HOPS is understood as being in direct proportion to total angular momentum difference between two basis vectors \cite{10.1103/PhysRevA.91.023801}. However, this result relies on external parameters, such as the rotation angle of the optical element \cite{10.1364/OL.42.003447}, while the intrinsic Berry curvature on the HOPS has not been discussed.

In this work, we develop a theoretical framework for the intrinsic Berry curvature on the generalized Poincaré sphere, which extends the concept of higher-order geometric phase to a more general setting. The geometric phase can be simply expressed as the magnetic flux of an effective magnetic field through the region enclosed by the state evolution trajectory. In our framework, the underlying reason for the change of geometric phase is clarified that not in the mordification of the effective central monopole charge, rather in the change of integral surface over which the megnetic flux is evaluated.

\section{BERRY CONNECTION, BERRY CURVATURE AND CHERN NUMBER ON GENERALIZED PS}
\begin{figure}
\includegraphics[width=\columnwidth]{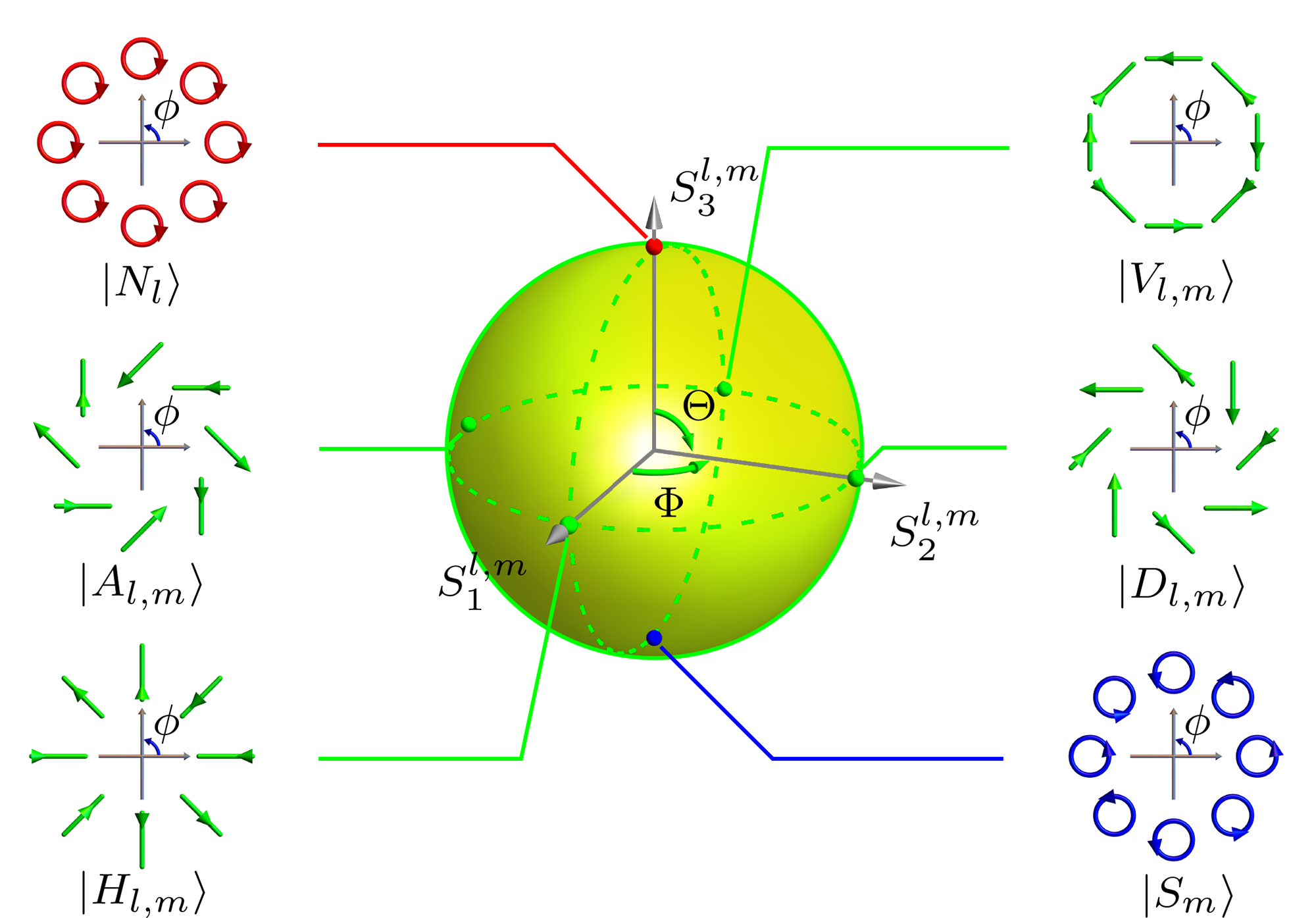}
\caption{Schematic illustration of the states on the generalized Poincaré sphere. Here, we assume $l = 0$ and $m = 2$. The north pole $|N_{l} \rangle$ denotes the RHC polarized plane-wave and $|S_{m} \rangle$ denotes the vortex light in LHC polarization. The points on the equator $|H_{l,m} \rangle$, $|V_{l,m} \rangle$, $|D_{l,m} \rangle$ and $|A_{l,m} \rangle$ denote different modes of vector light. The visualization of the Jones vector fields are arranged around the sphere, including the polarization ellipses at different positions, together with arrows indicating the instantaneous electric field evolution on the $x$-$y$ reference plane.}
\label{fig:Figure1}
\end{figure}

A representation of a class of structured light can be obtained by introducing vortex phase factor $e^{i\ell \phi}$ in left-handed circularly (LHC) and right-handed circularly (RHC) polarized basis vectors, where $\ell$ denotes topological charge, and $\phi$ is the azimuthal angle of the light beam on $x$-$y$ plane. The Jones vector \cite{Born1997} now can be regarded as a complex vector field being given by 
\begin{equation}
|\Psi_{l,m}\rangle = \cos \Big(\frac{\Theta} {2} \Big) e^{-i\Phi/2} |N_{l} \rangle + \sin \Big( \frac{\Theta} {2} \Big) e^{i\Phi/2} |S_{m} \rangle, 
\label{eq:Eq.(1)}
\end{equation}
where 
\begin{equation}
|N_{l} \rangle = ( \hat{\bm{\mathrm{e}}}_x + i \sigma_+ \hat{\bm{\mathrm{e}}}_y ) e^{il\phi}/\sqrt{2}, 
\label{eq:Eq.(2)}
\end{equation}
\begin{equation}
|S_{m} \rangle = ( \hat{\bm{\mathrm{e}}}_x + i \sigma_- \hat{\bm{\mathrm{e}}}_y ) e^{im\phi}/\sqrt{2}. 
\label{eq:Eq.(3)}
\end{equation}
Here, $\sigma_{\pm} = \pm 1$ is the helicity of the circularly polarized component, $l$ and $m$ are different topological charges. Equation (\ref{eq:Eq.(1)}) shows that the state is a combination of two new orthogonal basis states $|N_{l} \rangle$ and $|S_{m} \rangle$, where $\Theta$ and $\Phi$ determine the amplitude and phase difference of the normalized superposition coefficients, respectively. We map the state into a parameter space by introducing generalized Stokes paramters \cite{10.1103/PhysRevA.91.023801}
\begin{align}
S_0^{l,m} &= |\psi^l_N|^2 + |\psi^m_S|^2 \nonumber \\ 
&= 1, 
\label{eq:Eq.(4)}
\end{align}
\begin{align}
S_1^{l,m} &= 2 |\psi^l_N| |\psi^m_S| \cos [\arg (\psi^l_N) - \arg (\psi^m_S)] \nonumber \\ 
&= \sin \Theta \cos \Phi, 
\label{eq:Eq.(5)}
\end{align}
\begin{align}
S_2^{l,m} &= 2 |\psi^l_N| |\psi^m_S| \sin [\arg (\psi^l_N) - \arg (\psi^m_S)] \nonumber \\ 
&= \sin \Theta \sin \Phi, 
\label{eq:Eq.(6)}
\end{align}
\begin{align}
S_3^{l,m} &= |\psi^l_N|^2 - |\psi^m_S|^2 \nonumber \\ 
&= \cos \Theta. 
\label{eq:Eq.(7)}
\end{align}
Here, $\psi^l_N = \langle N_{l} |\Psi_{l,m}\rangle$ and $\psi^m_S = \langle S_{m} |\Psi_{l,m}\rangle$ are two components of the state. The point represented the state, of which the position vector can be written as $\bm{\mathrm{R}} = \hat{\bm{\mathrm{e}}}_1 S_1^{l,m} + \hat{\bm{\mathrm{e}}}_2 S_2^{l,m} + \hat{\bm{\mathrm{e}}}_3 S_3^{l,m}$ in Cartrsian coordinate frame, locates on the surface of the unit sphere. Figure \ref{fig:Figure1} shows the states on the sphere. 

\begin{figure}
\includegraphics[width=\columnwidth]{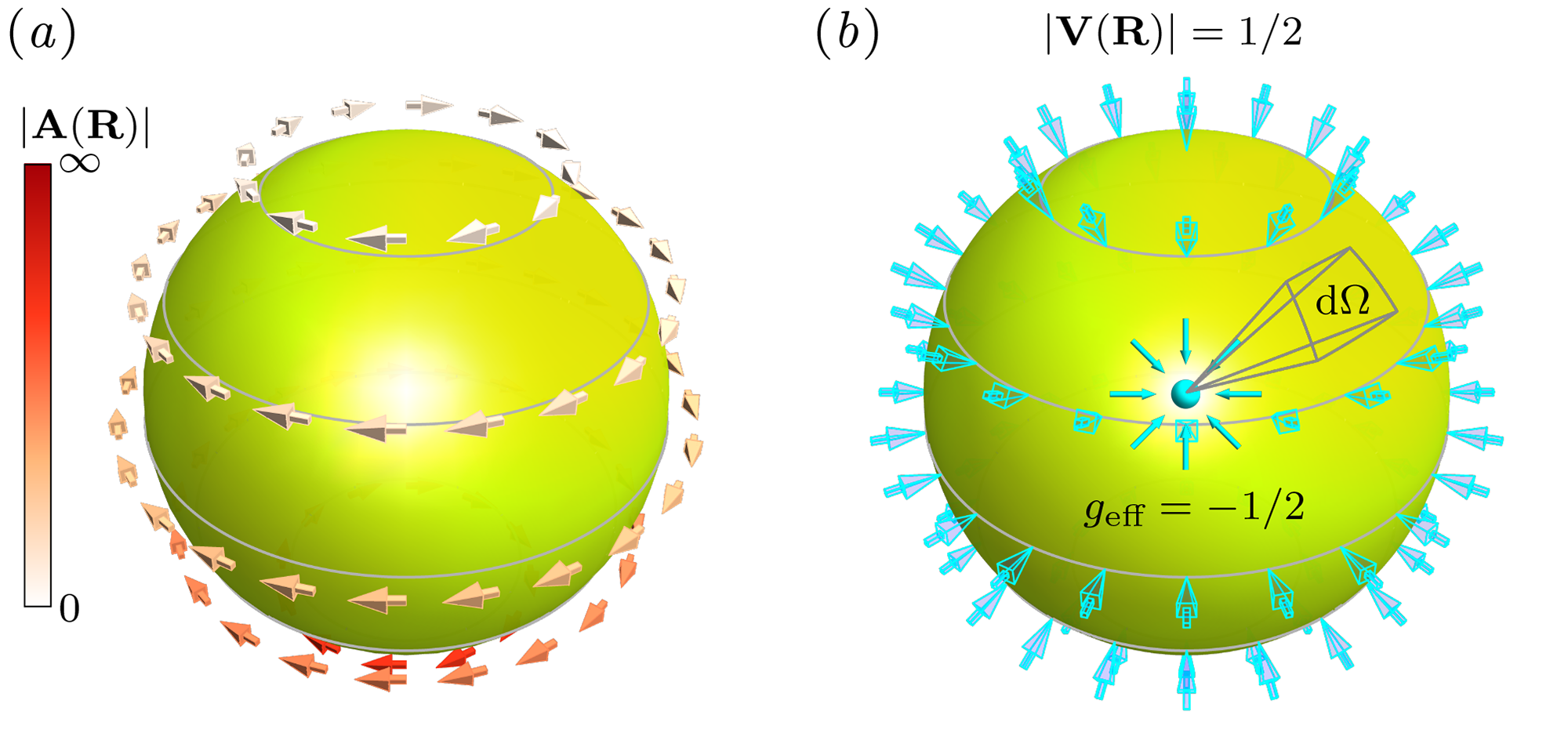}
\caption{(a) The Berry connection on the generalized Poincaré sphere. (b) Corresponding Berry curvature and effective magnetic monopole. Here, the arrow direction indicates the vector direction, while the color distribution represents its magnitude. Inset (a) shows that on the surface of the sphere, the Berry connection is directed along the latitude lines, behaving as a vector potential. Inset (b) shows that the direction of the Berry curvature is antiparallel to the normal to the sphere, similar to the magnetic field of a magnetic monopole located at the center of the sphere.}
\label{fig:Figure2}
\end{figure}

We now derive the Berry connection, Berry curvature, Chern number and geometric phase on the generalized Poincaré sphere. The Berry connection is given by 
\begin{equation}
\bm{\mathrm{A}} (\bm{\mathrm{R}}) = i \langle \Psi(\bm{\mathrm{R}}) | \nabla_{\bm{\mathrm{R}}}|\Psi(\bm{\mathrm{R}})\rangle.  
\label{eq:Eq.(8)}
\end{equation}
In sphereical coordinate frame, the position vector can be written as $\bm{\mathrm{R}} = \hat{\bm{\mathrm{e}}}_{\rho} \rho + \rho \ \hat{\bm{\mathrm{e}}}_{\Theta} \Theta + \rho \sin \Theta \ \hat{\bm{\mathrm{e}}}_{\Phi} \Phi$, and corresponding gradient operator is given by $\nabla_{\bm{\mathrm{R}}} = \frac{\partial} {\partial \rho} \hat{\bm{\mathrm{e}}}_{\rho} + \frac{1} {\rho}\frac{\partial} {\partial \Theta} \hat{\bm{\mathrm{e}}}_{\Theta} + \frac{1} {\rho \sin \Theta}\frac{\partial} {\partial \Phi} \hat{\bm{\mathrm{e}}}_{\Phi}$. We substitude Eq.(\ref{eq:Eq.(1)}) into Eq.(\ref{eq:Eq.(8)}), and get the Berry connction in Appendix \ref{app: APP.A} as 
\begin{equation}
\bm{\mathrm{A}} (\bm{\mathrm{R}}) = \frac{1} {\rho \sin \Theta} \frac{\cos \Theta - 1} {2} \hat{\bm{\mathrm{e}}}_{\Phi}. 
\label{eq:Eq.(9)}
\end{equation}
Here, since only the azimuthal component is nonzero, the vector potential is always directed along the lines of latitude on the sphere, as shown in Fig.\ref{fig:Figure2}. The characteristic behavior of the Berry curvature is consistent with that of the vector potential of a magnetic monopole \cite{10.1103/PhysRevB.72.035108}. Under our convention, its magnitude vanishes at the north pole of the sphere, diverges to positive infinity at the south pole.

The curl of the Berry connection is the Berry curvature, which plays a role as a magnatic field. It is given by 
\begin{equation}
\bm{\mathrm{V}} (\bm{\mathrm{R}}) = \nabla_{\bm{\mathrm{R}}} \times \bm{\mathrm{A}} (\bm{\mathrm{R}}). 
\label{eq:Eq.(10)}
\end{equation}
Substituding Eq.(\ref{eq:Eq.(9)}) into Eq.(\ref{eq:Eq.(10)}), we get 
\begin{equation}
\bm{\mathrm{V}} (\bm{\mathrm{R}}) = -\frac{1} {2\rho^2} \hat{\bm{\mathrm{e}}}_{\rho}. 
\label{eq:Eq.(11)}
\end{equation}
Eq.(\ref{eq:Eq.(11)}) shows that the effective magnetic field behaves as generated by a magnetic monopole with a constant charge located at the center of the sphere, which is is antiparallel to the normal to the sphere. The monopole charge $g_{\text{eff}} = -1/2$ can be obtained from Gauss's law for the field. Similarly, the Chern number is given by 
\begin{equation}
C = \frac{1} {2\pi} \iint_{\Sigma} \bm{\mathrm{V}} (\bm{\mathrm{R}}) \cdot \mathrm{d} \bm{\mathrm{S}}. 
\label{eq:Eq.(12)}
\end{equation}
Here, $\mathrm{d} \bm{\mathrm{S}} = \rho^2 \sin \Theta \mathrm{d} \rho \mathrm{d} \Theta \mathrm{d} \Phi \hat{\bm{\mathrm{e}}}_{\rho}$ is the surface element, and the integration is perfomed over the entire sphere. The resulting Chern number 
\begin{equation}
C = -1 
\label{eq:Eq.(13)}
\end{equation}
indicates a negtive unit topological charge associated with the Berry curvature. If we adopt the opposite phase convention to that in Eq.(\ref{eq:Eq.(1)}), that is, $|\Psi_{l,m}\rangle = \cos (\frac{\Theta} {2}) e^{i\Phi/2} |N_{l} \rangle + \sin (\frac{\Theta} {2}) e^{-i\Phi/2} |S_{m} \rangle$, we will obtain a positive Chern number $C = +1$.

The geometric phase arises in a continuous and cyclic state evolution in the parameter space $\bm{\mathrm{R}}$. Analogously to the magnetic flux, the geometric phase can be interpreted as the integration of the Berry curvature through the enclosed surface, which is given by 
\begin{equation}
\gamma_{\text{geo}} = \iint_{S} \bm{\mathrm{V}} (\bm{\mathrm{R}}) \cdot \mathrm{d} \bm{\mathrm{S}}. 
\label{eq:Eq.(14)}
\end{equation}
Here, $S$ is the area enclosed by the state evolution trajectory. Substituding Eq.(\ref{eq:Eq.(10)}) into Eq.(\ref{eq:Eq.(14)}), we get the geometric phase 
\begin{equation}
\gamma_{\text{geo}} = -\frac{1} {2} \Omega, 
\label{eq:Eq.(15)}
\end{equation}
where $\Omega$ is the solid angle enclosed by the state evolution trajectory.

Equations (\ref{eq:Eq.(9)}), (\ref{eq:Eq.(11)}), (\ref{eq:Eq.(13)}) and (\ref{eq:Eq.(15)}) show that the Berry connection, Berry curvature, Chern number and geometric phase do not explicitly involve the TAM. For the higher-order geometric phase, the association with the topological charge lies in the integration area of the magnetic flux, rather than in the magnetic field strength. When considering this phase, it is necessary to analyze the evolution of the state. The propagation of light through a q-plate\cite{10.1364/OE.14.004208, 10.1063/1.4878409, 10.1038/srep07840, 10.1103/PhysRevApplied.10.034061} provides a natural example of such a process, allowing us to apply Eq.(\ref{eq:Eq.(15)}) to calculate the geometric phase, which will be discussed in detail in the next section.

\section{STATE EVOLUTION AND GEOMETRIC PHASE ON GENERALIZED PS}
\begin{figure}
\includegraphics[width=\columnwidth]{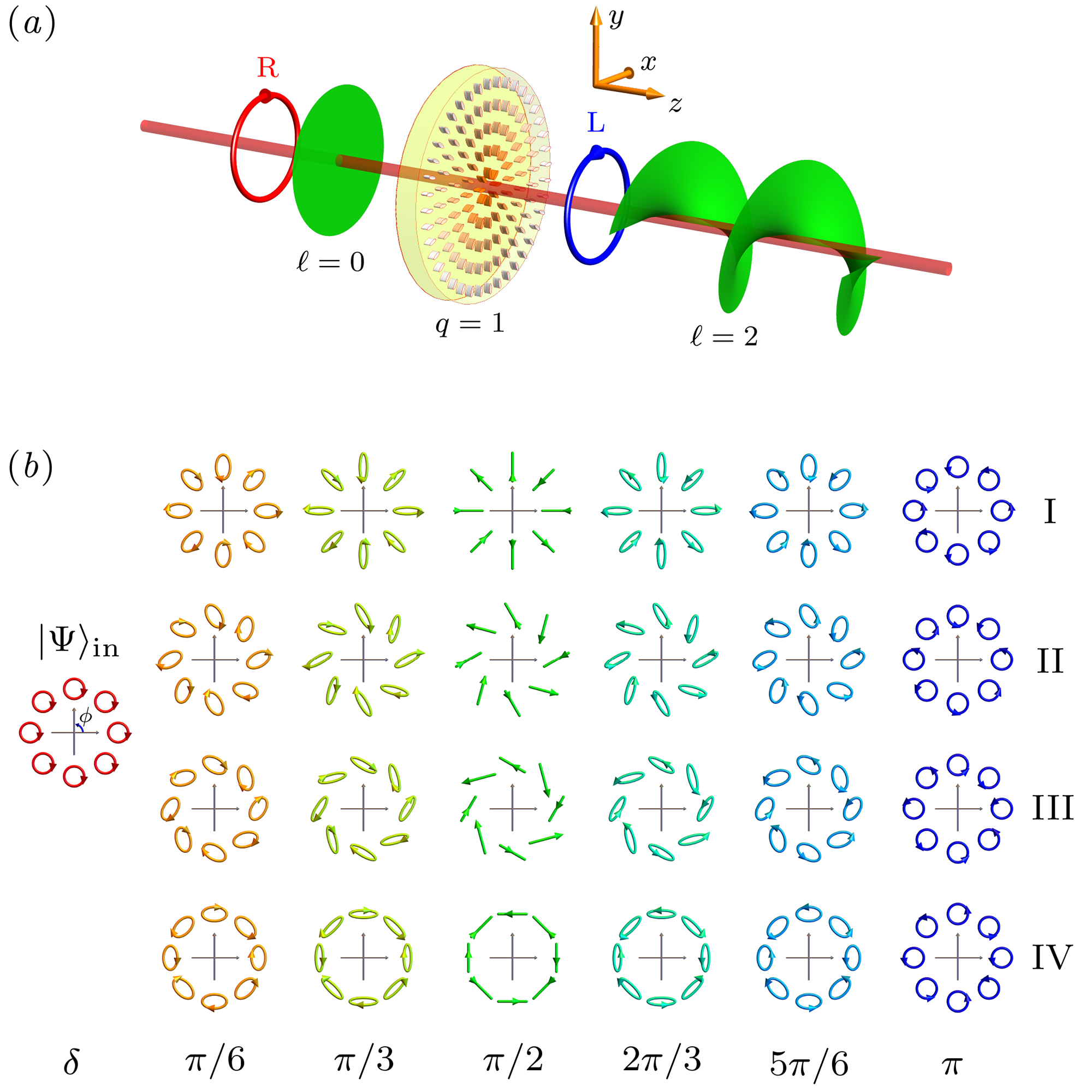}
\caption{(a) Schematic illustration of the propagation of a RHC polarized plane-wave through a q-plate. (b) Relevent state evolution. Here, we considered four distinct optical axis orientation distributions $\alpha_{\text{I-IV}}$ with the same topological charge $q = 1$. The phase retardation $\delta$ ranging from $0$ to $\pi$.}
\label{fig:Figure3}
\end{figure}

We now derive the trajectory on the sphere when a incident light in one of the TAM eigen states $|\Psi_{\text{in}}\rangle = |N_{l} \rangle$ passes through an inhomogeneous anisotropic media \cite{10.1103/PhysRevA.94.043851, PhysRevLett.96.163905}. Assuming the distrubution of the optical axis orientation of a q-plate can be written as 
\begin{equation}
\alpha (\phi) = q\phi + \alpha (0), 
\label{eq:Eq.(16)}
\end{equation}
where $\phi$ denotes azimuthal angle on $x$-$y$ plane, $q$ represents topological charge, $\alpha (0)$ is the local orientation angle of the optical axis at $\phi = 0$, the resulting light can be obtained in Appendix \ref{app: APP.B} as 
\begin{equation}
|\Psi_{\text{out}} (\delta)\rangle = \cos \Big(\frac{\delta} {2} \Big) |N_{l} \rangle + \sin \Big( \frac{\delta} {2} \Big) e^{i [2\alpha (0) - \pi/2]} |S_{m} \rangle. 
\label{eq:Eq.(17)}
\end{equation}
Here, $m = l+2q$, and $\delta$ denotes the phase retardation of the q-plate. An example of the state evolution is shown in Fig.\ref{fig:Figure3}. Equation (\ref{eq:Eq.(17)}) shows that the state evolutes along the longitude line with the azimuthal angle $2\alpha (0) - \pi/2$ from the north pole to the south pole on the generalized PS. When the phase retadation of the q-plate $\delta$ is equal to $\pi$, different longitude lines eventually converge at the south pole, with a final phase difference given by 
\begin{equation}
\gamma_{\text{fin}} = 2\Delta \alpha (0). 
\label{eq:Eq.(18)}
\end{equation}
Equation (\ref{eq:Eq.(18)}) shows that the geometric phase depends only on the local orientation angle of the optical axis $\alpha (0)$.

\begin{figure}
\includegraphics[width=\columnwidth]{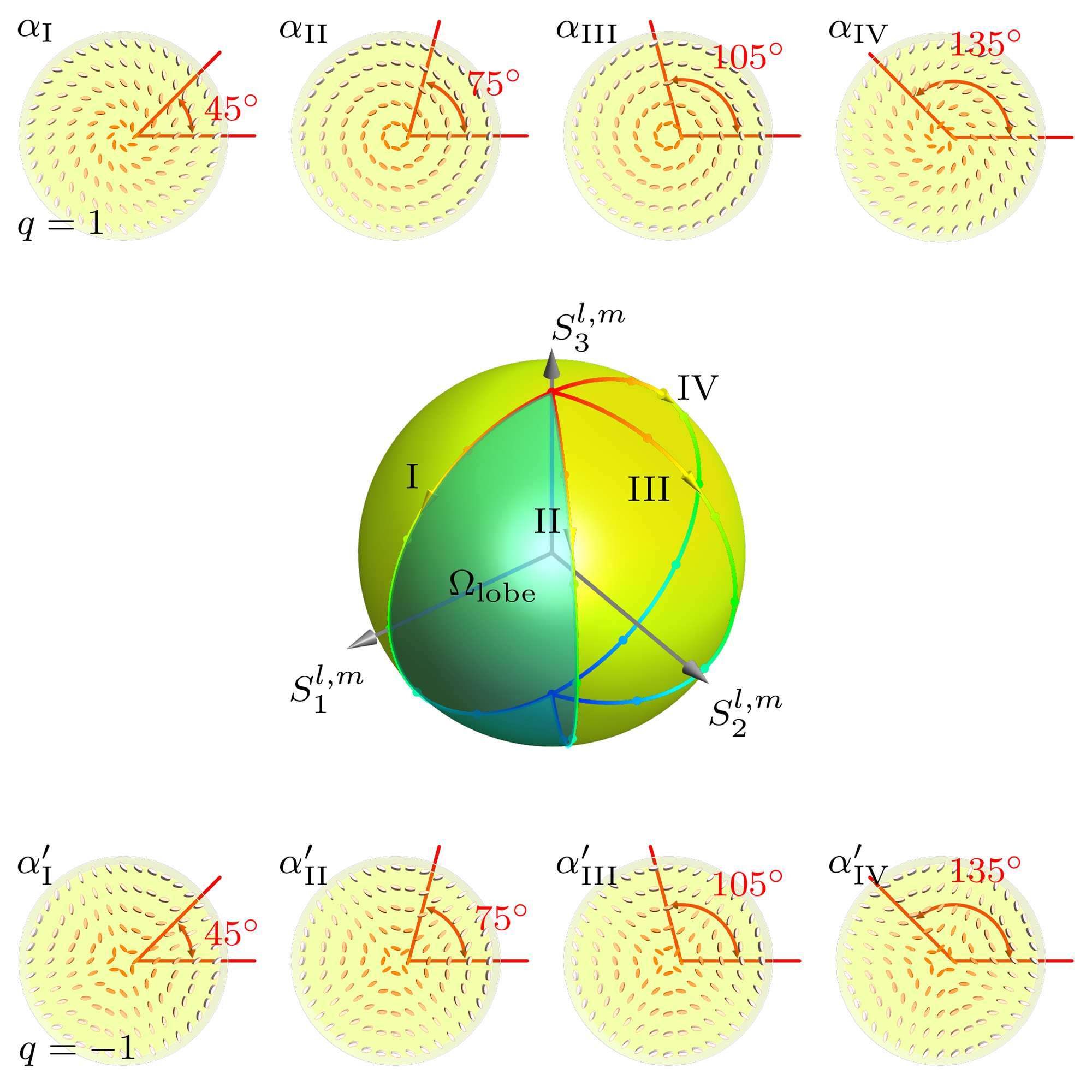}
\caption{Optical axis orientation distributions of the q-plates and related higher-order geometric phase on the generalized Poincaré sphere. Here, the trajectory of the state evolution is shown by color-graded longitude lines, and the geometric phase between the different longitude lines is the integration of the Berry curvature over the area subtended. We present an example with $q = -1$ additionally, combining with the case of $q = 1$, to show that the geometric phase depends only on the local orientation angle of the optical axis at $\phi = 0$.}
\label{fig:Figure4}
\end{figure}

Now we consider the flux of the Berry curvature on the sphere, as shown in Fig.\ref{fig:Figure4}. The two distinct longitude lines enclosed a lobe-shaped region, which subtend a solid angle 
\begin{equation}
\Omega_{\text{lobe}} = 4\Delta \alpha (0). 
\label{eq:Eq.(19)}
\end{equation}
Therefore, combining Eq.(\ref{eq:Eq.(18)}) and Eq.(\ref{eq:Eq.(19)}), the geometric representation of the phase difference can be written as 
\begin{equation}
\gamma_{\text{fin}} = \frac{1} {2}\Omega_{\text{lobe}}. 
\label{eq:Eq.(20)}
\end{equation}
Here, in comparison with Eq.(\ref{eq:Eq.(15)}), the result differs by a minus sign, which is reconciled by reversing the trajectory from the north pole to the south pole. The crucial point is that the factor $1/2$ in Eq.(\ref{eq:Eq.(20)}), being consisitent with the effective magnetic monopole charge $g_{\text{eff}}$, originates from the intrinsic Berry curvature on the generalized PS \cite{10.1364/JOSAA.538106}. 

\begin{figure}
\includegraphics[width=\columnwidth]{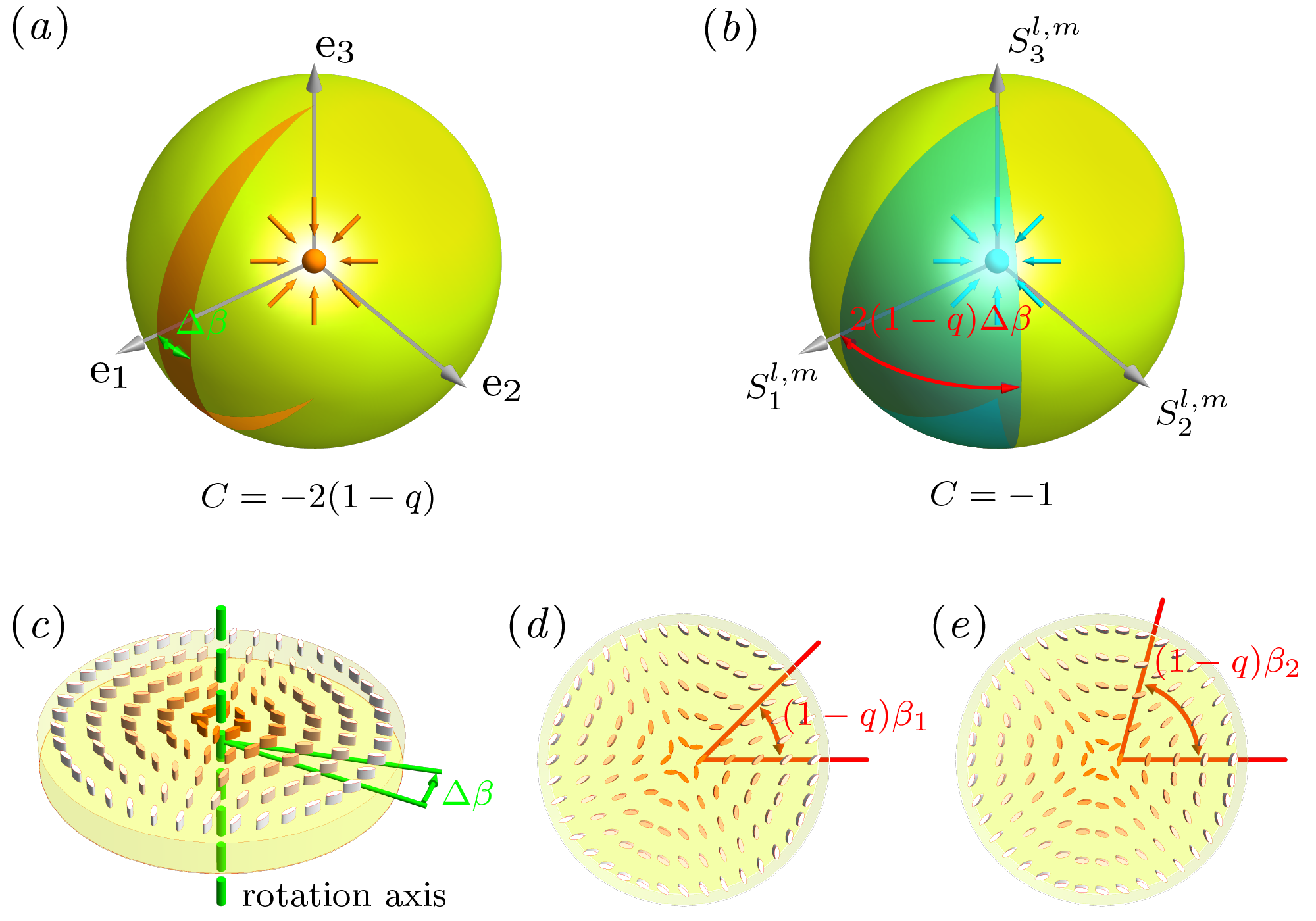}
\caption{Comparison between (a) the previous model and (b) the model we developed. (c) Schematic illustration of that the rotation of the q-plate changes by $\Delta \beta = \beta_2 - \beta_1$. The local optical axis angle of the q-plate $\alpha (0)$ before (d) and after (e) rotation. Here, two models give the same result of the geometric phase, while the geometric interperatations are different.}
\label{fig:Figure5}
\end{figure}

In typical experiments, the redistrubution of the optical axis orientation is attributed only to a rotation of the entire q-plate along $z$-axis. The rotation angle is assumed to be $\beta$, and the initial orientation angle of the optical axis $\alpha (0)$ is given by 
\begin{equation}
\alpha (0) = (1-q)\beta. 
\label{eq:Eq.(21)}
\end{equation}
Substituding Eq.(\ref{eq:Eq.(21)}) into Eq.(\ref{eq:Eq.(18)}), we immediately obtain 
\begin{equation}
\gamma_{\text{rot}} = 2(1-q)\Delta \beta. 
\label{eq:Eq.(22)}
\end{equation}
Since $\gamma_{\text{rot}}$ is directly propotional to $1-q$, it had been previously assumed that the effective magnetic monopole charge $g_{\text{eff}}$ is equal to the TAM difference between two basis states and infer that the solid angle $\Omega_{\text{lobe}}$ is equal to $2\Delta \beta$. However, in our framework, as demonstrated by our previous analysis, the longitude line locates at the positon with the azimuthal angle $2(1-q)\beta - \pi/2$, which means two different trajectories enclosed an solid angle given by 
\begin{equation}
\Omega_{\text{lobe}} = 4(1-q)\Delta \beta. 
\label{eq:Eq.(23)}
\end{equation}
In a word, the geometric phase is scaled by the difference in the TAM of the basis states, but this scaling arises from a change in Berry curvature integration area rather than from a change in the Berry curvature itself, as shown in Fig.\ref{fig:Figure5}. 

We revisit the phenomenon of zero geometric phase, which was previously attributed to zero Berry curvature \cite{10.1364/OL.42.003447}. When $q=1$, due to the rotational symmetry of the q-plate, the state always evolutes along an unchanged longitude line $2\alpha (0) - \pi/2$, which regardless the rotation angle $\beta$. Consequently, the enclosed area over which the Berry curvature integrates vanishes, despite the nonzero Berry curvature. In addition, the conclusion of the geometric phase being a flux of the intrinsic Berry curvature can also be extended to a more general condition. During spin-orbit interactions, the actual state evolution trajectory may deviate the longitude line, arising a geometric phase having no representation. Our theory may provide a fundamental frame to describe the geometric on the generalized PS for an arbitrary state evolution trajectory. 

\section{CONCLUSIONS}
In conclusion, we have developed a theoretical framework that establishes the independence of the Berry curvature from the topological charge of the generalized PS. Within this framework, the trajectory of the state evolution depends on the topological charge, and the geomeric phase is given by the integral of the Berry curvature over the area enclosed by the trajectory. This distinction provides a unified description of the relationship between state evolution and geometric phase, and offers a convenient approach to analyze higher-order geometric phase. Furthermore, the framework is applicable to both classical \cite{10.1103/PhysRevLett.61.19, 10.1103/PhysRevLett.81.4828, 10.1364/OE.25.011564, 10.1002/lpor.202000576, 10.1002/andp.202300250} and quantum systems\cite{10.1364/OE.19.009714, 10.1103/PhysRevA.88.033801, 10.1103/PhysRevA.62.022109, 10.1038/s41566-025-01795-x, 10.1098/rsta.2023.0327, 10.1103/PhysRevLett.122.013601, 10.29026/oes.2025.250006}. 

\begin{acknowledgments}
This work was supported by the National Natural Science Foundation of China (Grant No. 12674362) and Hunan Provincial Key Research and Development Program (Grant No. 2026QK3014).
\end{acknowledgments}

\appendix
\section{CALCULATION OF BERRY CONNECTION AND BERRY CURVATURE}
\label{app: APP.A}
In this appendix we give a detailed calculation of the Berry curvature of the generalized PS. In the sphereical coordinate frame, the Berry curvature can be written as 
\begin{equation}
A_{\rho} = i \langle \Psi (\bm{\mathrm{R}})| \frac{\partial} {\partial \rho} |\Psi (\bm{\mathrm{R}})\rangle, 
\label{eq:Eq.(A1)}
\end{equation}
\begin{equation}
A_{\Theta} = i \langle \Psi (\bm{\mathrm{R}})| \frac{1} {\rho} \frac{\partial} {\partial \Theta} |\Psi (\bm{\mathrm{R}})\rangle, 
\label{eq:Eq.(A2)}
\end{equation}
\begin{equation}
A_{\Phi} = i \langle \Psi (\bm{\mathrm{R}})| \frac{1} {\rho \sin \Theta} \frac{\partial} {\partial \Phi} |\Psi (\bm{\mathrm{R}})\rangle. 
\label{eq:Eq.(A3)}
\end{equation}
Since the state $|\Psi (\bm{\mathrm{R}})\rangle$ has been normalized, $\frac{\partial} {\partial \rho}|\Psi (\bm{\mathrm{R}})\rangle = 0$. Therefore, 
\begin{equation}
A_{\rho} = 0. 
\label{eq:Eq.(A4)}
\end{equation}
From Eq.(\ref{eq:Eq.(1)}), we know that 
\begin{align}
\frac{\partial} {\partial \Theta} |\Psi (\bm{\mathrm{R}})\rangle &= -\frac{1} {2} \sin \Big( \frac{\Theta} {2} \Big) e^{-i\Phi/2} |N_{l} \rangle \nonumber \\
&\quad + \frac{1} {2} \cos \Big(\frac{\Theta} {2} \Big) e^{i\Phi/2} |S_{m} \rangle. 
\label{eq:Eq.(A5)}
\end{align}
Substituding it into Eq.(\ref{eq:Eq.(A2)}), we get 
\begin{align}
A_{\Theta} &= -\frac{1} {2\rho} \sin \Big( \frac{\Theta} {2} \Big) \cos \Big( \frac{\Theta} {2} \Big) \langle N_{l} |N_{l} \rangle \nonumber \\
&\quad + \frac{1} {2\rho} \sin \Big( \frac{\Theta} {2} \Big) \cos \Big(\frac{\Theta} {2} \Big) \langle S_{m} |S_{m} \rangle \nonumber \\
&= 0. 
\label{eq:Eq.(A6)}
\end{align}
From Eq.(\ref{eq:Eq.(1)}), we get 
\begin{align}
\frac{\partial} {\partial \Phi} |\Psi (\bm{\mathrm{R}})\rangle &= -\frac{i} {2} \cos \Big(\frac{\Theta} {2} \Big) e^{-i\Phi/2} |N_{l} \rangle \nonumber \\
&\quad + \frac{i} {2} \sin \Big( \frac{\Theta} {2} \Big) e^{i\Phi/2} |S_{m} \rangle. 
\label{eq:Eq.(A7)}
\end{align}
Substituding Eq.(\ref{eq:Eq.(A7)}) into Eq.(\ref{eq:Eq.(A3)}), we get 
\begin{align}
A_{\Phi} &= \frac{1} {\rho \sin \Theta} \Big[ -\frac{i} {2} \cos^2 \Big( \frac{\Theta} {2} \Big) \langle N_{l} |N_{l} \rangle \nonumber \\
&\quad + \frac{i} {2} \sin^2 \Big( \frac{\Theta} {2} \Big) \langle S_{m} |S_{m} \rangle \Big] \nonumber \\
&= \frac{1} {\rho \sin \Theta} \frac{\cos \Theta - 1} {2}. 
\label{eq:Eq.(A8)}
\end{align}
The curl of the Berry curvature, is given by 
\begin{equation}
\bm{\mathrm{V}} = \frac{1} {\rho^2 \sin \Theta} \left |\begin{matrix} \hat{\bm{\mathrm{e}}}_{\rho} & \rho \ \hat{\bm{\mathrm{e}}}_{\Theta} & \rho \sin \Theta \ \hat{\bm{\mathrm{e}}}_{\Phi} \\ \frac{\partial} {\partial \rho} & \frac{\partial} {\partial \Theta} & \frac{\partial} {\partial \Phi} \\ A_{\rho} & \rho A_{\Theta} & \rho \sin \Theta A_{\Phi}\end{matrix} \right |. 
\label{eq:Eq.(A9)}
\end{equation}
Since $A_{\rho} = A_{\Theta} =0$, we can get 
\begin{equation}
V_{\Theta} = V_{\Phi} =0. 
\label{eq:Eq.(A10)}
\end{equation}
We substitude Eq.(\ref{eq:Eq.(A8)}) into Eq.(\ref{eq:Eq.(A9)}), and get 
\begin{align}
V_{\rho} &= \frac{1} {\rho^2 \sin \Theta} \frac{\partial} {\partial \Theta} (\rho \sin \Theta A_{\Phi}) \nonumber \\
&= -\frac{1} {2\rho^2}. 
\label{eq:Eq.(A11)}
\end{align}

\section{JONES MATRIX CALCULATIONS OF THE STATE EVOLUTION IN THE WAVE-PLATE WITH SPACIAL VARYING OPTICAL AXES}
\label{app: APP.B}
In this appendix we give a detailed calculation of state evolution in the q-plate. The Jones matrix is given by
\begin{align}
T &= M(\alpha) \begin{bmatrix} e^{-i \delta/2} & 0 \\ 0 & e^{i \delta/2} \end{bmatrix} M^{-1}(\alpha) \nonumber \\
&= \cos \Big( \frac {\delta} {2} \Big) \begin{bmatrix} 1 & 0 \\ 0 & 1 \end{bmatrix} \nonumber \\
&\quad - i \sin \Big( \frac {\delta} {2} \Big) \begin{bmatrix} \cos (2\alpha) & \sin (2\alpha) \\ \sin (2\alpha) & -\cos (2\alpha) \end{bmatrix}. 
\label{eq:Eq.(B1)}
\end{align}
Here, $M(\alpha)$ is SU(2) rotation matrix, given by 
\begin{equation}
M(\alpha) = \begin{bmatrix} \cos \alpha & -\sin \alpha \\ \sin \alpha & \cos \alpha\end{bmatrix}, 
\label{eq:Eq.(B2)}
\end{equation}
and $\alpha$ is the orientation angle of the local optical axis, given by Eq.(\ref{eq:Eq.(16)}). After the light in state $|\Psi_{\text{in}}\rangle$ passing through the q-plate, the resulting light can be written as 
\begin{align}
\begin{bmatrix} E_x \\ E_y \end{bmatrix}_{out} &= \cos \Big(\frac{\Theta} {2} \Big) \frac{1} {\sqrt{2}} \begin{bmatrix} 1 \\ i \sigma_+ \end{bmatrix} e^{il\phi}  \nonumber \\ 
& - i \sin \Big(\frac{\Theta} {2} \Big)\begin{bmatrix} \cos (2\alpha) & \sin (2\alpha) \\ \sin (2\alpha) & -\cos (2\alpha) \end{bmatrix} \frac{1} {\sqrt{2}}\begin{bmatrix} 1 \\ i \sigma_+ \end{bmatrix}  e^{il\phi} \nonumber \\
&= \cos \Big(\frac{\Theta} {2} \Big) \frac{1} {\sqrt{2}} \begin{bmatrix} 1 \\ i \sigma_+ \end{bmatrix} e^{il\phi} \nonumber \\
&\quad + \sin \Big(\frac{\Theta} {2} \Big) e^{i [2\alpha (0) - \pi/2]} \frac{1} {\sqrt{2}}\begin{bmatrix} 1 \\ i \sigma_- \end{bmatrix}  e^{i(l+2q)\phi}. 
\label{eq:Eq.(B3)}
\end{align}

\bibliographystyle{apsrev4-2}
\bibliography{references}

\begin{thebibliography}{49}%
\makeatletter
\providecommand \@ifxundefined [1]{%
 \@ifx{#1\undefined}
}%
\providecommand \@ifnum [1]{%
 \ifnum #1\expandafter \@firstoftwo
 \else \expandafter \@secondoftwo
 \fi
}%
\providecommand \@ifx [1]{%
 \ifx #1\expandafter \@firstoftwo
 \else \expandafter \@secondoftwo
 \fi
}%
\providecommand \natexlab [1]{#1}%
\providecommand \enquote  [1]{``#1''}%
\providecommand \bibnamefont  [1]{#1}%
\providecommand \bibfnamefont [1]{#1}%
\providecommand \citenamefont [1]{#1}%
\providecommand \href@noop [0]{\@secondoftwo}%
\providecommand \href [0]{\begingroup \@sanitize@url \@href}%
\providecommand \@href[1]{\@@startlink{#1}\@@href}%
\providecommand \@@href[1]{\endgroup#1\@@endlink}%
\providecommand \@sanitize@url [0]{\catcode `\\12\catcode `\$12\catcode
  `\&12\catcode `\#12\catcode `\^12\catcode `\_12\catcode `\%12\relax}%
\providecommand \@@startlink[1]{}%
\providecommand \@@endlink[0]{}%
\providecommand \url  [0]{\begingroup\@sanitize@url \@url }%
\providecommand \@url [1]{\endgroup\@href {#1}{\urlprefix }}%
\providecommand \urlprefix  [0]{URL }%
\providecommand \Eprint [0]{\href }%
\providecommand \doibase [0]{https://doi.org/}%
\providecommand \selectlanguage [0]{\@gobble}%
\providecommand \bibinfo  [0]{\@secondoftwo}%
\providecommand \bibfield  [0]{\@secondoftwo}%
\providecommand \translation [1]{[#1]}%
\providecommand \BibitemOpen [0]{}%
\providecommand \bibitemStop [0]{}%
\providecommand \bibitemNoStop [0]{.\EOS\space}%
\providecommand \EOS [0]{\spacefactor3000\relax}%
\providecommand \BibitemShut  [1]{\csname bibitem#1\endcsname}%
\let\auto@bib@innerbib\@empty
\bibitem [{\citenamefont {Berry}(1984)}]{10.1098/rspa.1984.0023}%
  \BibitemOpen
  \bibfield  {author} {\bibinfo {author} {\bibfnamefont {M.~V.}\ \bibnamefont
  {Berry}},\ }\href {https://doi.org/10.1098/rspa.1984.0023} {\bibfield
  {journal} {\bibinfo  {journal} {Proc. R. Soc. London A}\ }\textbf {\bibinfo
  {volume} {392}},\ \bibinfo {pages} {45} (\bibinfo {year} {1984})}\BibitemShut
  {NoStop}%
\bibitem [{\citenamefont {Aharonov}\ and\ \citenamefont
  {Anandan}(1987)}]{10.1103/PhysRevLett.58.1593}%
  \BibitemOpen
  \bibfield  {author} {\bibinfo {author} {\bibfnamefont {Y.}~\bibnamefont
  {Aharonov}}\ and\ \bibinfo {author} {\bibfnamefont {J.}~\bibnamefont
  {Anandan}},\ }\href {https://doi.org/10.1103/PhysRevLett.58.1593} {\bibfield
  {journal} {\bibinfo  {journal} {Phys. Rev. Lett.}\ }\textbf {\bibinfo
  {volume} {58}},\ \bibinfo {pages} {1593} (\bibinfo {year}
  {1987})}\BibitemShut {NoStop}%
\bibitem [{\citenamefont {Samuel}\ and\ \citenamefont
  {Bhandari}(1988)}]{10.1103/PhysRevLett.60.2339}%
  \BibitemOpen
  \bibfield  {author} {\bibinfo {author} {\bibfnamefont {J.}~\bibnamefont
  {Samuel}}\ and\ \bibinfo {author} {\bibfnamefont {R.}~\bibnamefont
  {Bhandari}},\ }\href {https://doi.org/10.1103/PhysRevLett.60.2339} {\bibfield
   {journal} {\bibinfo  {journal} {Phys. Rev. Lett.}\ }\textbf {\bibinfo
  {volume} {60}},\ \bibinfo {pages} {2339} (\bibinfo {year}
  {1988})}\BibitemShut {NoStop}%
\bibitem [{\citenamefont {Jisha}\ \emph {et~al.}(2021)\citenamefont {Jisha},
  \citenamefont {Nolte},\ and\ \citenamefont
  {Alberucci}}]{10.1002/lpor.202100003}%
  \BibitemOpen
  \bibfield  {author} {\bibinfo {author} {\bibfnamefont {C.~P.}\ \bibnamefont
  {Jisha}}, \bibinfo {author} {\bibfnamefont {S.}~\bibnamefont {Nolte}},\ and\
  \bibinfo {author} {\bibfnamefont {A.}~\bibnamefont {Alberucci}},\ }\href
  {https://doi.org/https://doi.org/10.1002/lpor.202100003} {\bibfield
  {journal} {\bibinfo  {journal} {Laser \& Photonics Rev.}\ }\textbf {\bibinfo
  {volume} {15}},\ \bibinfo {pages} {2100003} (\bibinfo {year}
  {2021})}\BibitemShut {NoStop}%
\bibitem [{\citenamefont {Poincaré}(1892)}]{Poincaré1892}%
  \BibitemOpen
  \bibfield  {author} {\bibinfo {author} {\bibfnamefont {H.}~\bibnamefont
  {Poincaré}},\ }\href@noop {} {\emph {\bibinfo {title} {Theorie Mathematique
  de la Lumiere}}}\ (\bibinfo  {publisher} {Gauthiers Villars},\ \bibinfo
  {address} {Paris},\ \bibinfo {year} {1892})\BibitemShut {NoStop}%
\bibitem [{\citenamefont {Jerrard}(1954)}]{10.1364/JOSA.44.000634}%
  \BibitemOpen
  \bibfield  {author} {\bibinfo {author} {\bibfnamefont {H.~G.}\ \bibnamefont
  {Jerrard}},\ }\href {https://doi.org/10.1364/JOSA.44.000634} {\bibfield
  {journal} {\bibinfo  {journal} {J. Opt. Soc. Am.}\ }\textbf {\bibinfo
  {volume} {44}},\ \bibinfo {pages} {634} (\bibinfo {year} {1954})}\BibitemShut
  {NoStop}%
\bibitem [{\citenamefont {Pancharatnam}(1956)}]{Pancharatnam1956}%
  \BibitemOpen
  \bibfield  {author} {\bibinfo {author} {\bibfnamefont {S.}~\bibnamefont
  {Pancharatnam}},\ }\href {https://doi.org/10.1007/BF03046095} {\bibfield
  {journal} {\bibinfo  {journal} {J. Mod. Opt.}\ }\textbf {\bibinfo {volume}
  {44}},\ \bibinfo {pages} {247} (\bibinfo {year} {1956})}\BibitemShut
  {NoStop}%
\bibitem [{\citenamefont {Berry}(1987)}]{10.1080/09500348714551321}%
  \BibitemOpen
  \bibfield  {author} {\bibinfo {author} {\bibfnamefont {M.~V.}\ \bibnamefont
  {Berry}},\ }\href {https://doi.org/10.1080/09500348714551321} {\bibfield
  {journal} {\bibinfo  {journal} {J. Mod. Opt.}\ }\textbf {\bibinfo {volume}
  {34}},\ \bibinfo {pages} {1401} (\bibinfo {year} {1987})}\BibitemShut
  {NoStop}%
\bibitem [{\citenamefont {Courtial}(1999)}]{10.1016/S0030-4018(99)00473-3}%
  \BibitemOpen
  \bibfield  {author} {\bibinfo {author} {\bibfnamefont {J.}~\bibnamefont
  {Courtial}},\ }\href
  {https://doi.org/https://doi.org/10.1016/S0030-4018(99)00473-3} {\bibfield
  {journal} {\bibinfo  {journal} {Opt. Commun.}\ }\textbf {\bibinfo {volume}
  {171}},\ \bibinfo {pages} {179} (\bibinfo {year} {1999})}\BibitemShut
  {NoStop}%
\bibitem [{\citenamefont {Garza-Soto}\ \emph {et~al.}(2023)\citenamefont
  {Garza-Soto}, \citenamefont {Hagen},\ and\ \citenamefont
  {Lopez-Mago}}]{10.1364/JOSAA.485485}%
  \BibitemOpen
  \bibfield  {author} {\bibinfo {author} {\bibfnamefont {L.}~\bibnamefont
  {Garza-Soto}}, \bibinfo {author} {\bibfnamefont {N.}~\bibnamefont {Hagen}},\
  and\ \bibinfo {author} {\bibfnamefont {D.}~\bibnamefont {Lopez-Mago}},\
  }\href {https://doi.org/10.1364/JOSAA.485485} {\bibfield  {journal} {\bibinfo
   {journal} {J. Opt. Soc. Am. A}\ }\textbf {\bibinfo {volume} {40}},\ \bibinfo
  {pages} {925} (\bibinfo {year} {2023})}\BibitemShut {NoStop}%
\bibitem [{\citenamefont {Bhandari}(1997)}]{10.1016/S0370-1573(96)00029-4}%
  \BibitemOpen
  \bibfield  {author} {\bibinfo {author} {\bibfnamefont {R.}~\bibnamefont
  {Bhandari}},\ }\href
  {https://doi.org/https://doi.org/10.1016/S0370-1573(96)00029-4} {\bibfield
  {journal} {\bibinfo  {journal} {Phys. Rep.}\ }\textbf {\bibinfo {volume}
  {281}},\ \bibinfo {pages} {1} (\bibinfo {year} {1997})}\BibitemShut {NoStop}%
\bibitem [{\citenamefont {Padgett}\ and\ \citenamefont
  {Courtial}(1999)}]{10.1364/OL.24.000430}%
  \BibitemOpen
  \bibfield  {author} {\bibinfo {author} {\bibfnamefont {M.~J.}\ \bibnamefont
  {Padgett}}\ and\ \bibinfo {author} {\bibfnamefont {J.}~\bibnamefont
  {Courtial}},\ }\href {https://doi.org/10.1364/OL.24.000430} {\bibfield
  {journal} {\bibinfo  {journal} {Opt. Lett.}\ }\textbf {\bibinfo {volume}
  {24}},\ \bibinfo {pages} {430} (\bibinfo {year} {1999})}\BibitemShut
  {NoStop}%
\bibitem [{\citenamefont {Galvez}\ \emph {et~al.}(2003)\citenamefont {Galvez},
  \citenamefont {Crawford}, \citenamefont {Sztul}, \citenamefont {Pysher},
  \citenamefont {Haglin},\ and\ \citenamefont
  {Williams}}]{10.1103/PhysRevLett.90.203901}%
  \BibitemOpen
  \bibfield  {author} {\bibinfo {author} {\bibfnamefont {E.~J.}\ \bibnamefont
  {Galvez}}, \bibinfo {author} {\bibfnamefont {P.~R.}\ \bibnamefont
  {Crawford}}, \bibinfo {author} {\bibfnamefont {H.~I.}\ \bibnamefont {Sztul}},
  \bibinfo {author} {\bibfnamefont {M.~J.}\ \bibnamefont {Pysher}}, \bibinfo
  {author} {\bibfnamefont {P.~J.}\ \bibnamefont {Haglin}},\ and\ \bibinfo
  {author} {\bibfnamefont {R.~E.}\ \bibnamefont {Williams}},\ }\href
  {https://doi.org/10.1103/PhysRevLett.90.203901} {\bibfield  {journal}
  {\bibinfo  {journal} {Phys. Rev. Lett.}\ }\textbf {\bibinfo {volume} {90}},\
  \bibinfo {pages} {203901} (\bibinfo {year} {2003})}\BibitemShut {NoStop}%
\bibitem [{\citenamefont {Bliokh}(2006)}]{10.1103/PhysRevLett.97.043901}%
  \BibitemOpen
  \bibfield  {author} {\bibinfo {author} {\bibfnamefont {K.~Y.}\ \bibnamefont
  {Bliokh}},\ }\href {https://doi.org/10.1103/PhysRevLett.97.043901} {\bibfield
   {journal} {\bibinfo  {journal} {Phys. Rev. Lett.}\ }\textbf {\bibinfo
  {volume} {97}},\ \bibinfo {pages} {043901} (\bibinfo {year}
  {2006})}\BibitemShut {NoStop}%
\bibitem [{\citenamefont {Allen}\ \emph {et~al.}(1992)\citenamefont {Allen},
  \citenamefont {Beijersbergen}, \citenamefont {Spreeuw},\ and\ \citenamefont
  {Woerdman}}]{10.1103/PhysRevA.45.8185}%
  \BibitemOpen
  \bibfield  {author} {\bibinfo {author} {\bibfnamefont {L.}~\bibnamefont
  {Allen}}, \bibinfo {author} {\bibfnamefont {M.~W.}\ \bibnamefont
  {Beijersbergen}}, \bibinfo {author} {\bibfnamefont {R.~J.~C.}\ \bibnamefont
  {Spreeuw}},\ and\ \bibinfo {author} {\bibfnamefont {J.~P.}\ \bibnamefont
  {Woerdman}},\ }\href {https://doi.org/10.1103/PhysRevA.45.8185} {\bibfield
  {journal} {\bibinfo  {journal} {Phys. Rev. A}\ }\textbf {\bibinfo {volume}
  {45}},\ \bibinfo {pages} {8185} (\bibinfo {year} {1992})}\BibitemShut
  {NoStop}%
\bibitem [{\citenamefont {Rubinsztein-Dunlop}\ \emph
  {et~al.}(2016)\citenamefont {Rubinsztein-Dunlop}, \citenamefont {Forbes},
  \citenamefont {Berry}, \citenamefont {Dennis}, \citenamefont {Andrews},
  \citenamefont {Mansuripur}, \citenamefont {Denz}, \citenamefont {Alpmann},
  \citenamefont {Banzer}, \citenamefont {Bauer}, \citenamefont {Karimi},
  \citenamefont {Marrucci}, \citenamefont {Padgett}, \citenamefont
  {Ritsch-Marte}, \citenamefont {Litchinitser}, \citenamefont {Bigelow},
  \citenamefont {Rosales-Guzmán}, \citenamefont {Belmonte}, \citenamefont
  {Torres}, \citenamefont {Neely}, \citenamefont {Baker}, \citenamefont
  {Gordon}, \citenamefont {Stilgoe}, \citenamefont {Romero}, \citenamefont
  {White}, \citenamefont {Fickler}, \citenamefont {Willner}, \citenamefont
  {Xie}, \citenamefont {McMorran},\ and\ \citenamefont
  {Weiner}}]{10.1088/2040-8978/19/1/013001}%
  \BibitemOpen
  \bibfield  {author} {\bibinfo {author} {\bibfnamefont {H.}~\bibnamefont
  {Rubinsztein-Dunlop}}, \bibinfo {author} {\bibfnamefont {A.}~\bibnamefont
  {Forbes}}, \bibinfo {author} {\bibfnamefont {M.~V.}\ \bibnamefont {Berry}},
  \bibinfo {author} {\bibfnamefont {M.~R.}\ \bibnamefont {Dennis}}, \bibinfo
  {author} {\bibfnamefont {D.~L.}\ \bibnamefont {Andrews}}, \bibinfo {author}
  {\bibfnamefont {M.}~\bibnamefont {Mansuripur}}, \bibinfo {author}
  {\bibfnamefont {C.}~\bibnamefont {Denz}}, \bibinfo {author} {\bibfnamefont
  {C.}~\bibnamefont {Alpmann}}, \bibinfo {author} {\bibfnamefont
  {P.}~\bibnamefont {Banzer}}, \bibinfo {author} {\bibfnamefont
  {T.}~\bibnamefont {Bauer}}, \bibinfo {author} {\bibfnamefont
  {E.}~\bibnamefont {Karimi}}, \bibinfo {author} {\bibfnamefont
  {L.}~\bibnamefont {Marrucci}}, \bibinfo {author} {\bibfnamefont
  {M.}~\bibnamefont {Padgett}}, \bibinfo {author} {\bibfnamefont
  {M.}~\bibnamefont {Ritsch-Marte}}, \bibinfo {author} {\bibfnamefont {N.~M.}\
  \bibnamefont {Litchinitser}}, \bibinfo {author} {\bibfnamefont {N.~P.}\
  \bibnamefont {Bigelow}}, \bibinfo {author} {\bibfnamefont {C.}~\bibnamefont
  {Rosales-Guzmán}}, \bibinfo {author} {\bibfnamefont {A.}~\bibnamefont
  {Belmonte}}, \bibinfo {author} {\bibfnamefont {J.~P.}\ \bibnamefont
  {Torres}}, \bibinfo {author} {\bibfnamefont {T.~W.}\ \bibnamefont {Neely}},
  \bibinfo {author} {\bibfnamefont {M.}~\bibnamefont {Baker}}, \bibinfo
  {author} {\bibfnamefont {R.}~\bibnamefont {Gordon}}, \bibinfo {author}
  {\bibfnamefont {A.~B.}\ \bibnamefont {Stilgoe}}, \bibinfo {author}
  {\bibfnamefont {J.}~\bibnamefont {Romero}}, \bibinfo {author} {\bibfnamefont
  {A.~G.}\ \bibnamefont {White}}, \bibinfo {author} {\bibfnamefont
  {R.}~\bibnamefont {Fickler}}, \bibinfo {author} {\bibfnamefont {A.~E.}\
  \bibnamefont {Willner}}, \bibinfo {author} {\bibfnamefont {G.}~\bibnamefont
  {Xie}}, \bibinfo {author} {\bibfnamefont {B.}~\bibnamefont {McMorran}},\ and\
  \bibinfo {author} {\bibfnamefont {A.~M.}\ \bibnamefont {Weiner}},\ }\href
  {https://doi.org/10.1088/2040-8978/19/1/013001} {\bibfield  {journal}
  {\bibinfo  {journal} {Journal of Optics}\ }\textbf {\bibinfo {volume} {19}},\
  \bibinfo {pages} {013001} (\bibinfo {year} {2016})}\BibitemShut {NoStop}%
\bibitem [{\citenamefont {Zhan}(2009)}]{10.1364/AOP.1.000001}%
  \BibitemOpen
  \bibfield  {author} {\bibinfo {author} {\bibfnamefont {Q.}~\bibnamefont
  {Zhan}},\ }\href {https://doi.org/10.1364/AOP.1.000001} {\bibfield  {journal}
  {\bibinfo  {journal} {Adv. Opt. Photonics}\ }\textbf {\bibinfo {volume}
  {1}},\ \bibinfo {pages} {1} (\bibinfo {year} {2009})}\BibitemShut {NoStop}%
\bibitem [{\citenamefont {Chen}\ \emph {et~al.}(2021)\citenamefont {Chen},
  \citenamefont {Wan},\ and\ \citenamefont {Zhan}}]{10.1117/1.AP.3.6.064001}%
  \BibitemOpen
  \bibfield  {author} {\bibinfo {author} {\bibfnamefont {J.}~\bibnamefont
  {Chen}}, \bibinfo {author} {\bibfnamefont {C.}~\bibnamefont {Wan}},\ and\
  \bibinfo {author} {\bibfnamefont {Q.}~\bibnamefont {Zhan}},\ }\href
  {https://doi.org/10.1117/1.AP.3.6.064001} {\bibfield  {journal} {\bibinfo
  {journal} {Adv. Photonics}\ }\textbf {\bibinfo {volume} {3}},\ \bibinfo
  {pages} {064001} (\bibinfo {year} {2021})}\BibitemShut {NoStop}%
\bibitem [{\citenamefont {Wang}\ \emph {et~al.}(2022)\citenamefont {Wang},
  \citenamefont {Tu}, \citenamefont {He}, \citenamefont {Xia}, \citenamefont
  {Hou}, \citenamefont {Li},\ and\ \citenamefont {Wang}}]{10.1063/5.0097923}%
  \BibitemOpen
  \bibfield  {author} {\bibinfo {author} {\bibfnamefont {Q.}~\bibnamefont
  {Wang}}, \bibinfo {author} {\bibfnamefont {C.}~\bibnamefont {Tu}}, \bibinfo
  {author} {\bibfnamefont {H.}~\bibnamefont {He}}, \bibinfo {author}
  {\bibfnamefont {Z.}~\bibnamefont {Xia}}, \bibinfo {author} {\bibfnamefont
  {X.}~\bibnamefont {Hou}}, \bibinfo {author} {\bibfnamefont {Y.}~\bibnamefont
  {Li}},\ and\ \bibinfo {author} {\bibfnamefont {H.}~\bibnamefont {Wang}},\
  }\href {https://doi.org/10.1063/5.0097923} {\bibfield  {journal} {\bibinfo
  {journal} {APL Photonics}\ }\textbf {\bibinfo {volume} {7}},\ \bibinfo
  {pages} {086102} (\bibinfo {year} {2022})}\BibitemShut {NoStop}%
\bibitem [{\citenamefont {Ren}\ \emph {et~al.}(2024)\citenamefont {Ren},
  \citenamefont {Cheng}, \citenamefont {Fan}, \citenamefont {Sun},
  \citenamefont {Zhu}, \citenamefont {Wan}, \citenamefont {Dong}, \citenamefont
  {Lou}, \citenamefont {Ding}, \citenamefont {Wang},\ and\ \citenamefont
  {Wang}}]{10.1063/5.0194981}%
  \BibitemOpen
  \bibfield  {author} {\bibinfo {author} {\bibfnamefont {Z.}~\bibnamefont
  {Ren}}, \bibinfo {author} {\bibfnamefont {Z.}~\bibnamefont {Cheng}}, \bibinfo
  {author} {\bibfnamefont {L.}~\bibnamefont {Fan}}, \bibinfo {author}
  {\bibfnamefont {R.}~\bibnamefont {Sun}}, \bibinfo {author} {\bibfnamefont
  {W.}~\bibnamefont {Zhu}}, \bibinfo {author} {\bibfnamefont {P.}~\bibnamefont
  {Wan}}, \bibinfo {author} {\bibfnamefont {B.}~\bibnamefont {Dong}}, \bibinfo
  {author} {\bibfnamefont {Y.}~\bibnamefont {Lou}}, \bibinfo {author}
  {\bibfnamefont {J.}~\bibnamefont {Ding}}, \bibinfo {author} {\bibfnamefont
  {X.}~\bibnamefont {Wang}},\ and\ \bibinfo {author} {\bibfnamefont
  {H.}~\bibnamefont {Wang}},\ }\href {https://doi.org/10.1063/5.0194981}
  {\bibfield  {journal} {\bibinfo  {journal} {APL Photonics}\ }\textbf
  {\bibinfo {volume} {9}},\ \bibinfo {pages} {031304} (\bibinfo {year}
  {2024})}\BibitemShut {NoStop}%
\bibitem [{\citenamefont {Shen}\ \emph {et~al.}(2020)\citenamefont {Shen},
  \citenamefont {Wang}, \citenamefont {Fu}, \citenamefont {Naidoo},\ and\
  \citenamefont {Forbes}}]{10.1103/PhysRevA.102.031501}%
  \BibitemOpen
  \bibfield  {author} {\bibinfo {author} {\bibfnamefont {Y.}~\bibnamefont
  {Shen}}, \bibinfo {author} {\bibfnamefont {Z.}~\bibnamefont {Wang}}, \bibinfo
  {author} {\bibfnamefont {X.}~\bibnamefont {Fu}}, \bibinfo {author}
  {\bibfnamefont {D.}~\bibnamefont {Naidoo}},\ and\ \bibinfo {author}
  {\bibfnamefont {A.}~\bibnamefont {Forbes}},\ }\href
  {https://doi.org/10.1103/PhysRevA.102.031501} {\bibfield  {journal} {\bibinfo
   {journal} {Phys. Rev. A}\ }\textbf {\bibinfo {volume} {102}},\ \bibinfo
  {pages} {031501(R)} (\bibinfo {year} {2020})}\BibitemShut {NoStop}%
\bibitem [{\citenamefont {Milione}\ \emph {et~al.}(2011)\citenamefont
  {Milione}, \citenamefont {Sztul}, \citenamefont {Nolan},\ and\ \citenamefont
  {Alfano}}]{10.1103/PhysRevLett.107.053601}%
  \BibitemOpen
  \bibfield  {author} {\bibinfo {author} {\bibfnamefont {G.}~\bibnamefont
  {Milione}}, \bibinfo {author} {\bibfnamefont {H.~I.}\ \bibnamefont {Sztul}},
  \bibinfo {author} {\bibfnamefont {D.~A.}\ \bibnamefont {Nolan}},\ and\
  \bibinfo {author} {\bibfnamefont {R.~R.}\ \bibnamefont {Alfano}},\ }\href
  {https://doi.org/10.1103/PhysRevLett.107.053601} {\bibfield  {journal}
  {\bibinfo  {journal} {Phys. Rev. Lett.}\ }\textbf {\bibinfo {volume} {107}},\
  \bibinfo {pages} {053601} (\bibinfo {year} {2011})}\BibitemShut {NoStop}%
\bibitem [{\citenamefont {Chen}\ \emph {et~al.}(2014)\citenamefont {Chen},
  \citenamefont {Zhou}, \citenamefont {Liu}, \citenamefont {Ling},
  \citenamefont {Luo},\ and\ \citenamefont {Wen}}]{10.1364/OL.39.005274}%
  \BibitemOpen
  \bibfield  {author} {\bibinfo {author} {\bibfnamefont {S.}~\bibnamefont
  {Chen}}, \bibinfo {author} {\bibfnamefont {X.}~\bibnamefont {Zhou}}, \bibinfo
  {author} {\bibfnamefont {Y.}~\bibnamefont {Liu}}, \bibinfo {author}
  {\bibfnamefont {X.}~\bibnamefont {Ling}}, \bibinfo {author} {\bibfnamefont
  {H.}~\bibnamefont {Luo}},\ and\ \bibinfo {author} {\bibfnamefont
  {S.}~\bibnamefont {Wen}},\ }\href {https://doi.org/10.1364/OL.39.005274}
  {\bibfield  {journal} {\bibinfo  {journal} {Opt. Lett.}\ }\textbf {\bibinfo
  {volume} {39}},\ \bibinfo {pages} {5274} (\bibinfo {year}
  {2014})}\BibitemShut {NoStop}%
\bibitem [{\citenamefont {Ren}\ \emph {et~al.}(2015)\citenamefont {Ren},
  \citenamefont {Kong}, \citenamefont {Li}, \citenamefont {Qian}, \citenamefont
  {Li}, \citenamefont {Tu},\ and\ \citenamefont {Wang}}]{10.1364/OE.23.026586}%
  \BibitemOpen
  \bibfield  {author} {\bibinfo {author} {\bibfnamefont {Z.}~\bibnamefont
  {Ren}}, \bibinfo {author} {\bibfnamefont {L.}~\bibnamefont {Kong}}, \bibinfo
  {author} {\bibfnamefont {S.}~\bibnamefont {Li}}, \bibinfo {author}
  {\bibfnamefont {S.}~\bibnamefont {Qian}}, \bibinfo {author} {\bibfnamefont
  {Y.}~\bibnamefont {Li}}, \bibinfo {author} {\bibfnamefont {C.}~\bibnamefont
  {Tu}},\ and\ \bibinfo {author} {\bibfnamefont {H.}~\bibnamefont {Wang}},\
  }\href {https://doi.org/10.1364/OE.23.026586} {\bibfield  {journal} {\bibinfo
   {journal} {Opt. Express}\ }\textbf {\bibinfo {volume} {23}},\ \bibinfo
  {pages} {26586} (\bibinfo {year} {2015})}\BibitemShut {NoStop}%
\bibitem [{\citenamefont {Naidoo}\ \emph {et~al.}(2016)\citenamefont {Naidoo},
  \citenamefont {Roux}, \citenamefont {Dudley}, \citenamefont {Litvin},
  \citenamefont {Piccirillo}, \citenamefont {Marrucci},\ and\ \citenamefont
  {Forbes}}]{10.1038/nphoton.2016.37}%
  \BibitemOpen
  \bibfield  {author} {\bibinfo {author} {\bibfnamefont {D.}~\bibnamefont
  {Naidoo}}, \bibinfo {author} {\bibfnamefont {F.~S.}\ \bibnamefont {Roux}},
  \bibinfo {author} {\bibfnamefont {A.}~\bibnamefont {Dudley}}, \bibinfo
  {author} {\bibfnamefont {I.}~\bibnamefont {Litvin}}, \bibinfo {author}
  {\bibfnamefont {B.}~\bibnamefont {Piccirillo}}, \bibinfo {author}
  {\bibfnamefont {L.}~\bibnamefont {Marrucci}},\ and\ \bibinfo {author}
  {\bibfnamefont {A.}~\bibnamefont {Forbes}},\ }\href
  {https://doi.org/10.1038/nphoton.2016.37} {\bibfield  {journal} {\bibinfo
  {journal} {Nat. Photonics}\ }\textbf {\bibinfo {volume} {10}},\ \bibinfo
  {pages} {327} (\bibinfo {year} {2016})}\BibitemShut {NoStop}%
\bibitem [{\citenamefont {Milione}\ \emph {et~al.}(2012)\citenamefont
  {Milione}, \citenamefont {Evans}, \citenamefont {Nolan},\ and\ \citenamefont
  {Alfano}}]{10.1103/PhysRevLett.108.190401}%
  \BibitemOpen
  \bibfield  {author} {\bibinfo {author} {\bibfnamefont {G.}~\bibnamefont
  {Milione}}, \bibinfo {author} {\bibfnamefont {S.}~\bibnamefont {Evans}},
  \bibinfo {author} {\bibfnamefont {D.~A.}\ \bibnamefont {Nolan}},\ and\
  \bibinfo {author} {\bibfnamefont {R.~R.}\ \bibnamefont {Alfano}},\ }\href
  {https://doi.org/10.1103/PhysRevLett.108.190401} {\bibfield  {journal}
  {\bibinfo  {journal} {Phys. Rev. Lett.}\ }\textbf {\bibinfo {volume} {108}},\
  \bibinfo {pages} {190401} (\bibinfo {year} {2012})}\BibitemShut {NoStop}%
\bibitem [{\citenamefont {Yi}\ \emph {et~al.}(2015)\citenamefont {Yi},
  \citenamefont {Liu}, \citenamefont {Ling}, \citenamefont {Zhou},
  \citenamefont {Ke}, \citenamefont {Luo}, \citenamefont {Wen},\ and\
  \citenamefont {Fan}}]{10.1103/PhysRevA.91.023801}%
  \BibitemOpen
  \bibfield  {author} {\bibinfo {author} {\bibfnamefont {X.}~\bibnamefont
  {Yi}}, \bibinfo {author} {\bibfnamefont {Y.}~\bibnamefont {Liu}}, \bibinfo
  {author} {\bibfnamefont {X.}~\bibnamefont {Ling}}, \bibinfo {author}
  {\bibfnamefont {X.}~\bibnamefont {Zhou}}, \bibinfo {author} {\bibfnamefont
  {Y.}~\bibnamefont {Ke}}, \bibinfo {author} {\bibfnamefont {H.}~\bibnamefont
  {Luo}}, \bibinfo {author} {\bibfnamefont {S.}~\bibnamefont {Wen}},\ and\
  \bibinfo {author} {\bibfnamefont {D.}~\bibnamefont {Fan}},\ }\href
  {https://doi.org/10.1103/PhysRevA.91.023801} {\bibfield  {journal} {\bibinfo
  {journal} {Phys. Rev. A}\ }\textbf {\bibinfo {volume} {91}},\ \bibinfo
  {pages} {023801} (\bibinfo {year} {2015})}\BibitemShut {NoStop}%
\bibitem [{\citenamefont {Liu}\ \emph {et~al.}(2017{\natexlab{a}})\citenamefont
  {Liu}, \citenamefont {Liu}, \citenamefont {Zhou}, \citenamefont {Ling},
  \citenamefont {Shu}, \citenamefont {Luo},\ and\ \citenamefont
  {Wen}}]{10.1364/OL.42.003447}%
  \BibitemOpen
  \bibfield  {author} {\bibinfo {author} {\bibfnamefont {Y.}~\bibnamefont
  {Liu}}, \bibinfo {author} {\bibfnamefont {Z.}~\bibnamefont {Liu}}, \bibinfo
  {author} {\bibfnamefont {J.}~\bibnamefont {Zhou}}, \bibinfo {author}
  {\bibfnamefont {X.}~\bibnamefont {Ling}}, \bibinfo {author} {\bibfnamefont
  {W.}~\bibnamefont {Shu}}, \bibinfo {author} {\bibfnamefont {H.}~\bibnamefont
  {Luo}},\ and\ \bibinfo {author} {\bibfnamefont {S.}~\bibnamefont {Wen}},\
  }\href {https://doi.org/10.1364/OL.42.003447} {\bibfield  {journal} {\bibinfo
   {journal} {Opt. Lett.}\ }\textbf {\bibinfo {volume} {42}},\ \bibinfo {pages}
  {3447} (\bibinfo {year} {2017}{\natexlab{a}})}\BibitemShut {NoStop}%
\bibitem [{\citenamefont {Born}\ and\ \citenamefont {Wolf}(1997)}]{Born1997}%
  \BibitemOpen
  \bibfield  {author} {\bibinfo {author} {\bibfnamefont {M.}~\bibnamefont
  {Born}}\ and\ \bibinfo {author} {\bibfnamefont {E.}~\bibnamefont {Wolf}},\
  }\href@noop {} {\emph {\bibinfo {title} {Principles of Optics}}}\ (\bibinfo
  {publisher} {Cambridge University Press},\ \bibinfo {address} {Cambridge},\
  \bibinfo {year} {1997})\BibitemShut {NoStop}%
\bibitem [{\citenamefont {Bliokh}\ and\ \citenamefont
  {Freilikher}(2005)}]{10.1103/PhysRevB.72.035108}%
  \BibitemOpen
  \bibfield  {author} {\bibinfo {author} {\bibfnamefont {K.~Y.}\ \bibnamefont
  {Bliokh}}\ and\ \bibinfo {author} {\bibfnamefont {V.~D.}\ \bibnamefont
  {Freilikher}},\ }\href {https://doi.org/10.1103/PhysRevB.72.035108}
  {\bibfield  {journal} {\bibinfo  {journal} {Phys. Rev. B}\ }\textbf {\bibinfo
  {volume} {72}},\ \bibinfo {pages} {035108} (\bibinfo {year}
  {2005})}\BibitemShut {NoStop}%
\bibitem [{\citenamefont {Niv}\ \emph {et~al.}(2006)\citenamefont {Niv},
  \citenamefont {Biener}, \citenamefont {Kleiner},\ and\ \citenamefont
  {Hasman}}]{10.1364/OE.14.004208}%
  \BibitemOpen
  \bibfield  {author} {\bibinfo {author} {\bibfnamefont {A.}~\bibnamefont
  {Niv}}, \bibinfo {author} {\bibfnamefont {G.}~\bibnamefont {Biener}},
  \bibinfo {author} {\bibfnamefont {V.}~\bibnamefont {Kleiner}},\ and\ \bibinfo
  {author} {\bibfnamefont {E.}~\bibnamefont {Hasman}},\ }\href
  {https://doi.org/10.1364/OE.14.004208} {\bibfield  {journal} {\bibinfo
  {journal} {Opt. Express}\ }\textbf {\bibinfo {volume} {14}},\ \bibinfo
  {pages} {4208} (\bibinfo {year} {2006})}\BibitemShut {NoStop}%
\bibitem [{\citenamefont {Liu}\ \emph {et~al.}(2014)\citenamefont {Liu},
  \citenamefont {Ling}, \citenamefont {Yi}, \citenamefont {Zhou}, \citenamefont
  {Luo},\ and\ \citenamefont {Wen}}]{10.1063/1.4878409}%
  \BibitemOpen
  \bibfield  {author} {\bibinfo {author} {\bibfnamefont {Y.}~\bibnamefont
  {Liu}}, \bibinfo {author} {\bibfnamefont {X.}~\bibnamefont {Ling}}, \bibinfo
  {author} {\bibfnamefont {X.}~\bibnamefont {Yi}}, \bibinfo {author}
  {\bibfnamefont {X.}~\bibnamefont {Zhou}}, \bibinfo {author} {\bibfnamefont
  {H.}~\bibnamefont {Luo}},\ and\ \bibinfo {author} {\bibfnamefont
  {S.}~\bibnamefont {Wen}},\ }\href {https://doi.org/10.1063/1.4878409}
  {\bibfield  {journal} {\bibinfo  {journal} {Appl. Phys. Lett.}\ }\textbf
  {\bibinfo {volume} {104}},\ \bibinfo {pages} {191110} (\bibinfo {year}
  {2014})}\BibitemShut {NoStop}%
\bibitem [{\citenamefont {D'Ambrosio}\ \emph {et~al.}(2015)\citenamefont
  {D'Ambrosio}, \citenamefont {Baccari}, \citenamefont {Slussarenko},
  \citenamefont {Marrucci},\ and\ \citenamefont
  {Sciarrino}}]{10.1038/srep07840}%
  \BibitemOpen
  \bibfield  {author} {\bibinfo {author} {\bibfnamefont {V.}~\bibnamefont
  {D'Ambrosio}}, \bibinfo {author} {\bibfnamefont {F.}~\bibnamefont {Baccari}},
  \bibinfo {author} {\bibfnamefont {S.}~\bibnamefont {Slussarenko}}, \bibinfo
  {author} {\bibfnamefont {L.}~\bibnamefont {Marrucci}},\ and\ \bibinfo
  {author} {\bibfnamefont {F.}~\bibnamefont {Sciarrino}},\ }\href
  {https://doi.org/10.1038/srep07840} {\bibfield  {journal} {\bibinfo
  {journal} {Sci. Rep.}\ }\textbf {\bibinfo {volume} {5}},\ \bibinfo {pages}
  {7840} (\bibinfo {year} {2015})}\BibitemShut {NoStop}%
\bibitem [{\citenamefont {Xu}\ \emph {et~al.}(2018)\citenamefont {Xu},
  \citenamefont {Chen}, \citenamefont {Tang}, \citenamefont {Duan},
  \citenamefont {Ge}, \citenamefont {Ma}, \citenamefont {Wu}, \citenamefont
  {Hu},\ and\ \citenamefont {Lu}}]{10.1103/PhysRevApplied.10.034061}%
  \BibitemOpen
  \bibfield  {author} {\bibinfo {author} {\bibfnamefont {R.}~\bibnamefont
  {Xu}}, \bibinfo {author} {\bibfnamefont {P.}~\bibnamefont {Chen}}, \bibinfo
  {author} {\bibfnamefont {J.}~\bibnamefont {Tang}}, \bibinfo {author}
  {\bibfnamefont {W.}~\bibnamefont {Duan}}, \bibinfo {author} {\bibfnamefont
  {S.}~\bibnamefont {Ge}}, \bibinfo {author} {\bibfnamefont {L.}~\bibnamefont
  {Ma}}, \bibinfo {author} {\bibfnamefont {R.}~\bibnamefont {Wu}}, \bibinfo
  {author} {\bibfnamefont {W.}~\bibnamefont {Hu}},\ and\ \bibinfo {author}
  {\bibfnamefont {Y.}~\bibnamefont {Lu}},\ }\href
  {https://doi.org/10.1103/PhysRevApplied.10.034061} {\bibfield  {journal}
  {\bibinfo  {journal} {Phys. Rev. Appl.}\ }\textbf {\bibinfo {volume} {10}},\
  \bibinfo {pages} {034061} (\bibinfo {year} {2018})}\BibitemShut {NoStop}%
\bibitem [{\citenamefont {Suzuki}\ \emph {et~al.}(2016)\citenamefont {Suzuki},
  \citenamefont {Yamane}, \citenamefont {Oka}, \citenamefont {Toda},\ and\
  \citenamefont {Morita}}]{10.1103/PhysRevA.94.043851}%
  \BibitemOpen
  \bibfield  {author} {\bibinfo {author} {\bibfnamefont {M.}~\bibnamefont
  {Suzuki}}, \bibinfo {author} {\bibfnamefont {K.}~\bibnamefont {Yamane}},
  \bibinfo {author} {\bibfnamefont {K.}~\bibnamefont {Oka}}, \bibinfo {author}
  {\bibfnamefont {Y.}~\bibnamefont {Toda}},\ and\ \bibinfo {author}
  {\bibfnamefont {R.}~\bibnamefont {Morita}},\ }\href
  {https://doi.org/10.1103/PhysRevA.94.043851} {\bibfield  {journal} {\bibinfo
  {journal} {Phys. Rev. A}\ }\textbf {\bibinfo {volume} {94}},\ \bibinfo
  {pages} {043851} (\bibinfo {year} {2016})}\BibitemShut {NoStop}%
\bibitem [{\citenamefont {Marrucci}\ \emph {et~al.}(2006)\citenamefont
  {Marrucci}, \citenamefont {Manzo},\ and\ \citenamefont
  {Paparo}}]{PhysRevLett.96.163905}%
  \BibitemOpen
  \bibfield  {author} {\bibinfo {author} {\bibfnamefont {L.}~\bibnamefont
  {Marrucci}}, \bibinfo {author} {\bibfnamefont {C.}~\bibnamefont {Manzo}},\
  and\ \bibinfo {author} {\bibfnamefont {D.}~\bibnamefont {Paparo}},\ }\href
  {https://doi.org/10.1103/PhysRevLett.96.163905} {\bibfield  {journal}
  {\bibinfo  {journal} {Phys. Rev. Lett.}\ }\textbf {\bibinfo {volume} {96}},\
  \bibinfo {pages} {163905} (\bibinfo {year} {2006})}\BibitemShut {NoStop}%
\bibitem [{\citenamefont {Hagen}\ and\ \citenamefont
  {Garza-Soto}(2024)}]{10.1364/JOSAA.538106}%
  \BibitemOpen
  \bibfield  {author} {\bibinfo {author} {\bibfnamefont {N.}~\bibnamefont
  {Hagen}}\ and\ \bibinfo {author} {\bibfnamefont {L.}~\bibnamefont
  {Garza-Soto}},\ }\href {https://doi.org/10.1364/JOSAA.538106} {\bibfield
  {journal} {\bibinfo  {journal} {J. Opt. Soc. Am. A}\ }\textbf {\bibinfo
  {volume} {41}},\ \bibinfo {pages} {2014} (\bibinfo {year}
  {2024})}\BibitemShut {NoStop}%
\bibitem [{\citenamefont {Simon}\ \emph {et~al.}(1988)\citenamefont {Simon},
  \citenamefont {Kimble},\ and\ \citenamefont
  {Sudarshan}}]{10.1103/PhysRevLett.61.19}%
  \BibitemOpen
  \bibfield  {author} {\bibinfo {author} {\bibfnamefont {R.}~\bibnamefont
  {Simon}}, \bibinfo {author} {\bibfnamefont {H.~J.}\ \bibnamefont {Kimble}},\
  and\ \bibinfo {author} {\bibfnamefont {E.~C.~G.}\ \bibnamefont {Sudarshan}},\
  }\href {https://doi.org/10.1103/PhysRevLett.61.19} {\bibfield  {journal}
  {\bibinfo  {journal} {Phys. Rev. Lett.}\ }\textbf {\bibinfo {volume} {61}},\
  \bibinfo {pages} {19} (\bibinfo {year} {1988})}\BibitemShut {NoStop}%
\bibitem [{\citenamefont {Courtial}\ \emph {et~al.}(1998)\citenamefont
  {Courtial}, \citenamefont {Robertson}, \citenamefont {Dholakia},
  \citenamefont {Allen},\ and\ \citenamefont
  {Padgett}}]{10.1103/PhysRevLett.81.4828}%
  \BibitemOpen
  \bibfield  {author} {\bibinfo {author} {\bibfnamefont {J.}~\bibnamefont
  {Courtial}}, \bibinfo {author} {\bibfnamefont {D.~A.}\ \bibnamefont
  {Robertson}}, \bibinfo {author} {\bibfnamefont {K.}~\bibnamefont {Dholakia}},
  \bibinfo {author} {\bibfnamefont {L.}~\bibnamefont {Allen}},\ and\ \bibinfo
  {author} {\bibfnamefont {M.~J.}\ \bibnamefont {Padgett}},\ }\href
  {https://doi.org/10.1103/PhysRevLett.81.4828} {\bibfield  {journal} {\bibinfo
   {journal} {Phys. Rev. Lett.}\ }\textbf {\bibinfo {volume} {81}},\ \bibinfo
  {pages} {4828} (\bibinfo {year} {1998})}\BibitemShut {NoStop}%
\bibitem [{\citenamefont {Liu}\ \emph {et~al.}(2017{\natexlab{b}})\citenamefont
  {Liu}, \citenamefont {Liu}, \citenamefont {Ke}, \citenamefont {Zhou},
  \citenamefont {Liu}, \citenamefont {Luo},\ and\ \citenamefont
  {Wen}}]{10.1364/OE.25.011564}%
  \BibitemOpen
  \bibfield  {author} {\bibinfo {author} {\bibfnamefont {Z.}~\bibnamefont
  {Liu}}, \bibinfo {author} {\bibfnamefont {Y.}~\bibnamefont {Liu}}, \bibinfo
  {author} {\bibfnamefont {Y.}~\bibnamefont {Ke}}, \bibinfo {author}
  {\bibfnamefont {J.}~\bibnamefont {Zhou}}, \bibinfo {author} {\bibfnamefont
  {Y.}~\bibnamefont {Liu}}, \bibinfo {author} {\bibfnamefont {H.}~\bibnamefont
  {Luo}},\ and\ \bibinfo {author} {\bibfnamefont {S.}~\bibnamefont {Wen}},\
  }\href {https://doi.org/10.1364/OE.25.011564} {\bibfield  {journal} {\bibinfo
   {journal} {Opt.Express}\ }\textbf {\bibinfo {volume} {25}},\ \bibinfo
  {pages} {11564} (\bibinfo {year} {2017}{\natexlab{b}})}\BibitemShut {NoStop}%
\bibitem [{\citenamefont {Yue}\ \emph {et~al.}(2021)\citenamefont {Yue},
  \citenamefont {Aadhi}, \citenamefont {Piccoli}, \citenamefont {Aglieri},
  \citenamefont {Macaluso}, \citenamefont {Toma}, \citenamefont {Morandotti},\
  and\ \citenamefont {Razzari}}]{10.1002/lpor.202000576}%
  \BibitemOpen
  \bibfield  {author} {\bibinfo {author} {\bibfnamefont {F.}~\bibnamefont
  {Yue}}, \bibinfo {author} {\bibfnamefont {A.}~\bibnamefont {Aadhi}}, \bibinfo
  {author} {\bibfnamefont {R.}~\bibnamefont {Piccoli}}, \bibinfo {author}
  {\bibfnamefont {V.}~\bibnamefont {Aglieri}}, \bibinfo {author} {\bibfnamefont
  {R.}~\bibnamefont {Macaluso}}, \bibinfo {author} {\bibfnamefont
  {A.}~\bibnamefont {Toma}}, \bibinfo {author} {\bibfnamefont {R.}~\bibnamefont
  {Morandotti}},\ and\ \bibinfo {author} {\bibfnamefont {L.}~\bibnamefont
  {Razzari}},\ }\href {https://doi.org/https://doi.org/10.1002/lpor.202000576}
  {\bibfield  {journal} {\bibinfo  {journal} {Laser \& Photonics Rev.}\
  }\textbf {\bibinfo {volume} {15}},\ \bibinfo {pages} {2000576} (\bibinfo
  {year} {2021})}\BibitemShut {NoStop}%
\bibitem [{\citenamefont {Emile}\ and\ \citenamefont
  {Emile}(2023)}]{10.1002/andp.202300250}%
  \BibitemOpen
  \bibfield  {author} {\bibinfo {author} {\bibfnamefont {O.}~\bibnamefont
  {Emile}}\ and\ \bibinfo {author} {\bibfnamefont {J.}~\bibnamefont {Emile}},\
  }\href {https://doi.org/https://doi.org/10.1002/andp.202300250} {\bibfield
  {journal} {\bibinfo  {journal} {Ann. Phys.}\ }\textbf {\bibinfo {volume}
  {535}},\ \bibinfo {pages} {2300250} (\bibinfo {year} {2023})}\BibitemShut
  {NoStop}%
\bibitem [{\citenamefont {Holleczek}\ \emph {et~al.}(2011)\citenamefont
  {Holleczek}, \citenamefont {Aiello}, \citenamefont {Gabriel}, \citenamefont
  {Marquardt},\ and\ \citenamefont {Leuchs}}]{10.1364/OE.19.009714}%
  \BibitemOpen
  \bibfield  {author} {\bibinfo {author} {\bibfnamefont {A.}~\bibnamefont
  {Holleczek}}, \bibinfo {author} {\bibfnamefont {A.}~\bibnamefont {Aiello}},
  \bibinfo {author} {\bibfnamefont {C.}~\bibnamefont {Gabriel}}, \bibinfo
  {author} {\bibfnamefont {C.}~\bibnamefont {Marquardt}},\ and\ \bibinfo
  {author} {\bibfnamefont {G.}~\bibnamefont {Leuchs}},\ }\href
  {https://doi.org/10.1364/OE.19.009714} {\bibfield  {journal} {\bibinfo
  {journal} {Opt. Express}\ }\textbf {\bibinfo {volume} {19}},\ \bibinfo
  {pages} {9714} (\bibinfo {year} {2011})}\BibitemShut {NoStop}%
\bibitem [{\citenamefont {Kuratsuji}(2013)}]{10.1103/PhysRevA.88.033801}%
  \BibitemOpen
  \bibfield  {author} {\bibinfo {author} {\bibfnamefont {H.}~\bibnamefont
  {Kuratsuji}},\ }\href {https://doi.org/10.1103/PhysRevA.88.033801} {\bibfield
   {journal} {\bibinfo  {journal} {Phys. Rev. A}\ }\textbf {\bibinfo {volume}
  {88}},\ \bibinfo {pages} {033801} (\bibinfo {year} {2013})}\BibitemShut
  {NoStop}%
\bibitem [{\citenamefont {Sj{\"o}qvist}(2000)}]{10.1103/PhysRevA.62.022109}%
  \BibitemOpen
  \bibfield  {author} {\bibinfo {author} {\bibfnamefont {E.}~\bibnamefont
  {Sj{\"o}qvist}},\ }\href {https://doi.org/10.1103/PhysRevA.62.022109}
  {\bibfield  {journal} {\bibinfo  {journal} {Phys. Rev. A}\ }\textbf {\bibinfo
  {volume} {62}},\ \bibinfo {pages} {022109} (\bibinfo {year}
  {2000})}\BibitemShut {NoStop}%
\bibitem [{\citenamefont {Forbes}\ \emph {et~al.}(2025)\citenamefont {Forbes},
  \citenamefont {Nothlawala},\ and\ \citenamefont
  {Vallés}}]{10.1038/s41566-025-01795-x}%
  \BibitemOpen
  \bibfield  {author} {\bibinfo {author} {\bibfnamefont {A.}~\bibnamefont
  {Forbes}}, \bibinfo {author} {\bibfnamefont {F.}~\bibnamefont {Nothlawala}},\
  and\ \bibinfo {author} {\bibfnamefont {A.}~\bibnamefont {Vallés}},\ }\href
  {https://doi.org/10.1038/s41566-025-01795-x} {\bibfield  {journal} {\bibinfo
  {journal} {Nat. Photonics}\ }\textbf {\bibinfo {volume} {19}},\ \bibinfo
  {pages} {1291} (\bibinfo {year} {2025})}\BibitemShut {NoStop}%
\bibitem [{\citenamefont {Padgett}(2024)}]{10.1098/rsta.2023.0327}%
  \BibitemOpen
  \bibfield  {author} {\bibinfo {author} {\bibfnamefont {M.}~\bibnamefont
  {Padgett}},\ }\href {https://doi.org/10.1098/rsta.2023.0327} {\bibfield
  {journal} {\bibinfo  {journal} {Phil. Trans. R. Soc. A}\ }\textbf {\bibinfo
  {volume} {382}},\ \bibinfo {pages} {20230327} (\bibinfo {year}
  {2024})}\BibitemShut {NoStop}%
\bibitem [{\citenamefont {D'Ambrosio}\ \emph {et~al.}(2019)\citenamefont
  {D'Ambrosio}, \citenamefont {Carvacho}, \citenamefont {Agresti},
  \citenamefont {Marrucci},\ and\ \citenamefont
  {Sciarrino}}]{10.1103/PhysRevLett.122.013601}%
  \BibitemOpen
  \bibfield  {author} {\bibinfo {author} {\bibfnamefont {V.}~\bibnamefont
  {D'Ambrosio}}, \bibinfo {author} {\bibfnamefont {G.}~\bibnamefont
  {Carvacho}}, \bibinfo {author} {\bibfnamefont {I.}~\bibnamefont {Agresti}},
  \bibinfo {author} {\bibfnamefont {L.}~\bibnamefont {Marrucci}},\ and\
  \bibinfo {author} {\bibfnamefont {F.}~\bibnamefont {Sciarrino}},\ }\href
  {https://doi.org/10.1103/PhysRevLett.122.013601} {\bibfield  {journal}
  {\bibinfo  {journal} {Phys. Rev. Lett.}\ }\textbf {\bibinfo {volume} {122}},\
  \bibinfo {pages} {013601} (\bibinfo {year} {2019})}\BibitemShut {NoStop}%
\bibitem [{\citenamefont {Wang}\ \emph {et~al.}(2025)\citenamefont {Wang},
  \citenamefont {Shou}, \citenamefont {Liu}, \citenamefont {Yang},
  \citenamefont {Chen}, \citenamefont {Shu}, \citenamefont {Wen},\ and\
  \citenamefont {Luo}}]{10.29026/oes.2025.250006}%
  \BibitemOpen
  \bibfield  {author} {\bibinfo {author} {\bibfnamefont {Y.}~\bibnamefont
  {Wang}}, \bibinfo {author} {\bibfnamefont {Y.}~\bibnamefont {Shou}}, \bibinfo
  {author} {\bibfnamefont {J.}~\bibnamefont {Liu}}, \bibinfo {author}
  {\bibfnamefont {Q.}~\bibnamefont {Yang}}, \bibinfo {author} {\bibfnamefont
  {S.}~\bibnamefont {Chen}}, \bibinfo {author} {\bibfnamefont {W.}~\bibnamefont
  {Shu}}, \bibinfo {author} {\bibfnamefont {S.}~\bibnamefont {Wen}},\ and\
  \bibinfo {author} {\bibfnamefont {H.}~\bibnamefont {Luo}},\ }\href
  {https://doi.org/10.29026/oes.2025.250006} {\bibfield  {journal} {\bibinfo
  {journal} {Opto-Electronic Sci.}\ }\textbf {\bibinfo {volume} {4}},\ \bibinfo
  {pages} {250006} (\bibinfo {year} {2025})}\BibitemShut {NoStop}%
\end{thebibliography}%

\end{document}